\documentclass[preprint,showpacs,preprintnumbers,eqsecnum]{revtex4}

\usepackage{graphicx}
\usepackage{dcolumn}
\usepackage{bm}

\def\lesssim{\ \raise.3ex\hbox{$<$}\kern-0.8em\lower.7ex\hbox{$\sim$}\ }
\def\gesim{\ \raise.3ex\hbox{$>$}\kern-0.8em\lower.7ex\hbox{$\sim$}\ }
\font\scripti=cmmi7
\font\scriptscripti=cmmi5
\def\sib#1{\setbox0 = \hbox{\scripti #1}
  \kern-.02em\copy0\kern-\wd0
  \kern.04em\box0} 
\def\ssib#1{\setbox0 = \hbox{\scriptscripti #1}
  \kern-.02em\copy0\kern-\wd0
  \kern.04em\box0} 

\font\tenib=cmmib10 
\skewchar\tenib='177 \skewchar\tenib='177 \skewchar\tenib='177
\def\pbold#1{\setbox0 = \hbox{$ #1 $}
  \kern-.022em\copy0\kern-\wd0
  \kern.011em\copy0\kern-\wd0
  \kern.011em\copy0\kern-\wd0
  \kern.011em\copy0\kern-\wd0
  \kern.011em\box0} 

\begin{document}
\title{Kohn's theorem in a
  superfluid Fermi gas with a Feshbach resonance}
\author{Y. Ohashi}
\affiliation{Institute of Physics, University of Tsukuba, Tsukuba,
  Ibaraki 305, Japan}
\date{\today}
\begin{abstract}
We investigate the dipole mode in a superfluid gas of
Fermi atoms trapped in a harmonic potential. According to Kohn's
theorem, the frequency of this 
collective mode is not affected by an interaction
between the atoms and is always equal to the trap frequency. This
remarkable property, however, does not necessarily hold in an approximate
theory. 
We explicitly prove that the Hartree-Fock-Bogoliubov generalized
random phase approximation (HFB-GRPA), including a coupling
between fluctuations in the density and Cooper channels, is consistent
with both Kohn's theorem as well as Goldstone's theorem. 
This proof can be immediately extended to the
strong-coupling superfluid theory developed by Nozi\'eres 
and Schmitt-Rink (NSR),
where the effect of superfluid fluctuations is included
within the Gaussian level. As a result, the NSR-GRPA formalism can be
used to study collective modes in the BCS-BEC crossover region
in a manner which is consistent with Kohn's theorem.
We also include the effect of a Feshbach resonance and 
a condensate of the 
associated
molecular bound states. 
A detailed discussion is given of the unusual nature of the
Kohn mode eigenfunctions in a Fermi superfluid, in the
presence and absence of a Feshbach resonance.
When the molecular bosons feel a different trap
frequency from the Fermi atoms, the dipole frequency 
is shown to {\it depend} on
the strength of effective interaction associated with the Feshbach
resonance. 
\end{abstract}
\pacs{03.75.Ss, 03.75.Kk, 03.75.Nt}
\maketitle
%
\section{Introduction}
Among various collective excitations in an atomic gas trapped in a harmonic
potential, the dipole mode
has the unique property that its frequency is always equal to the
trap frequency, irrespective of the interaction between
the atoms. This remarkable property was originally proved by Kohn in the
context of the cyclotron frequency of electrons in metals\cite{Kohn}, 
and later it
was extended to an excitation spectrum of electrons in a quantum well 
produced in Al$_x$Ga$_{1-x}$As\cite{Brey,Dobson}. Recently, Kohn's theorem has 
been extensively discussed in the context of
Bose-Einstein condensation (BEC) of trapped atomic gases\cite{Pethick}.
\par
Kohn's theorem is a direct consequence of the translational invariance of 
particle-particle 
interaction term, $U_{\rm int}\equiv \sum_{i<j}u({\bf r}_i-{\bf r}_j)$. 
To show this, we consider an $N$-body system described by the Hamiltonian
\begin{equation}
H=\sum_{i=1}^N{{\bf p}_i^2 \over 2m}+U_{\rm int}
+{1 \over 2}m\sum_{i=1}^N
[\Omega_x^2x_i^2+\Omega_y^2y_i^2+\Omega_z^2z_i^2].
\label{eq.1.1}
\end{equation}
It is easy to see that the operators 
${\hat P}_\alpha\equiv\sum_{i=1}^N
[m\Omega_\alpha{\hat r}_{\alpha,i}-
i{\hat p}_{\alpha,i}]~(\alpha=x,y,z)$ satisfy the
commutation relations $[H,{\hat P}_\alpha]=\Omega_\alpha
{\hat P}_\alpha$\cite{Brey}. We note that the fact that 
$[{\hat P}_\alpha,U_{\rm int}]=0$ is
crucial to obtain this relation. If $|\Psi_0\rangle$ is the ground
state wave-function with energy $E_0$, the three excited states generated by
${\hat P}_\alpha$, 
$|\Psi_1\rangle_\alpha\equiv 
{\hat P}_\alpha|\Psi_0\rangle$ ($\alpha=x,y,z$),
are eigenstates with energies $E_\alpha=\Omega_\alpha+E_0$, 
with $U_{\rm int}$ having no effect. These excited states 
are referred to collectively as the ``Kohn modes"
in the theoretical literature
(the ``sloshing modes'' in the recent cold atom experimental
literature).
\par
While Kohn's theorem is exact, it is not a trivial problem as to whether
or not it holds when the interaction is
treated approximately. For example, in the mean-field approximation
($U_{\rm int}\to U_{\rm MF}=\sum_i\int d{\bf r}'u({\bf r}_i-
{\bf r}')n({\bf r}')$, where $n({\bf r}')$ is the particle density), the
translational invariance of the original (exact) $U_{\rm int}$ is
broken, which leads to $[{\hat P}_\alpha,U_{\rm MF}]\ne 0$. Since, in most
cases, we cannot treat many-body effects exactly, it is an important
 goal to obtain an approximation consistent
with Kohn's theorem. The nature of the Kohn mode has been extensively
studied in trapped Bose
gases\cite{Pethick,Stoof,Griffin2,Griffin,Fetter,Reidl}. 
In particular, Fetter
et. al.\cite{Fetter} and Reidl et. al.\cite{Reidl} proved that, 
in trapped Bose gases, the
Hartree-Fock random phase approximation (HF-RPA) is consistent with
this theorem at $T=0$ and at $T>0$, respectively. 
\par
This problem is particularly crucial for fermion superfluidity of atoms in a
trap, which is a topic of great current interest,
because the BCS pairing {\it approximation}
($U_{\rm BCS}\equiv
-U\Psi^\dagger_\uparrow({\bf r})\Psi^\dagger_\downarrow
({\bf r})\Psi_\downarrow({\bf r})\Psi_\uparrow({\bf r})\to\Delta
({\bf r})[\Psi^\dagger_\uparrow({\bf r})\Psi^\dagger_\downarrow
({\bf r})+h.c.]$, where $\Psi_\sigma({\bf r})$ is a fermion field operator
and $\Delta({\bf r})$ is the Cooper-pair order parameter) 
is almost always used in the study of the BCS superfluid phase. 
In a trap, the ${\bf r}$-dependent
order parameter $\Delta({\bf r})$ 
destroys the translational invariance of the pairing interaction. 
Besides this, in a trapped gas of Fermi atoms,
we need to take into account a Feshbach resonance and 
the associated molecules\cite{Timmermans1,Timmermans2}. The
enhanced pairing interaction mediated by the molecules
can be  
used to study superfluidity\cite{Holland1,Holland2,Kokkelmans1,Ohashi1,
Ohashi3,Ohashi4,Milstein}
in atomic Fermi gases, such as $^{40}$K and
$^6$Li\cite{Jin1,Ketterle1,Ohara,Jin4}.
In addition, since the strength of this effective interaction
is tunable by an external magnetic field, one can probe the BCS-BEC
crossover phenomenon\cite{Leggett,Nozieres,Randeria,Melo}, where 
the superfluidity continuously changes from a BCS-type to
a BEC of composite bosons\cite{Ohashi1,Ohashi3,Ohashi4,Milstein}.
Very recently, superfluidity both in the BEC
regime\cite{Jin2} and in the crossover
regime\cite{Jin10,Kinast,Bartenstein,Zwierlein} 
has been observed in $^{40}$K and $^6$Li. 
In dealing with such systems, one must ensure that 
approximate theories are
consistent with Kohn's theorem.
\par
In this paper, we give an explicit formal 
proof that the Hartree-Fock-Bogoliubov
generalized random phase approximation (HFB-GRPA), including a
coupling between density fluctuations with superfluid fluctuations, is
consistent with Kohn's theorem at all temperatures. 
We prove that this is also valid for the BCS-BEC crossover.
It is also valid for 
the case when one includes fluctuations around the 
mean-field approximation, such as first done by Nozi\`eres 
and Schmitt-Rink (NSR)\cite{Nozieres} at $T_{\rm c}$. This proof
can be easily extended to deal with systems with a Feshbach resonance.
Thus our results show that the NSR-GRPA formalism
can be safely used to study 
response functions and collective modes in the whole region of
BCS-BEC crossover, without fear of the breakdown of Kohn's theorem.
We use the same formalism to explicitly verify that a zero frequency
Goldstone sound mode arises associated with the Bose broken symmetry.
\par
We also show that, as expected, Kohn's theorem is no longer valid
in the presence of 
Feshbach resonance if the molecular bosons feel a different trap
frequency $\Omega_{\rm B}$ from the trap frequency $\Omega_F$ which
the Fermi
atoms feel. In this case,
the dipole mode frequency does {\it depend} on the strength of the
effective interaction as we go through the BCS-BEC
crossover regime. 
\par
We give the explicit forms for the Kohn mode eigenstates
in the BCS-BEC crossover region, without and with a Feshbach
resonance. In the case of a Feshbach resonance in the
crossover region, we show that the Kohn mode eigenstate is not simply
a rigid center of mass oscillation of the static condensate
profile. 
\par
This paper is organized as follows. In Sec. II, we prove that HFB-GRPA
is consistent with 
Kohn's theorem at all temperatures, in the absence of
the Feshbach resonance. The extension of this proof to the
NSR-GRPA treatment of fluctuations is also
discussed. In Sec. III, we generalize the proof to 
include the appearance of molecules associated with a
Feshbach resonance. We also consider the case when the 
Fermi atoms and quasi-molecules feel different trap frequencies.
Throughout this paper, we set $\hbar=1$ for
simplicity.
\par
\vskip10mm
\section{Kohn's theorem in the BCS approximation}
\par
\subsection{Hartree-Fock Bogoliubov approximation}
\vskip2mm
We consider a two-component Fermi gas trapped in a harmonic potential
described by the Hamiltonian,
\begin{eqnarray}
H
&=&
\sum_\sigma\int d{\bf r}
\Psi^\dagger_\sigma({\bf r})[-{\nabla^2 \over 2m}
+V_{\rm trap}-\mu]\Psi_\sigma({\bf r})
-
U\int d{\bf r}
\Psi_{\uparrow}^\dagger({\bf r})
\Psi_{\downarrow}^\dagger({\bf r})
\Psi_{\downarrow}({\bf r})
\Psi_{\uparrow}({\bf r}),
\label{eq.2.1}
\end{eqnarray}
where $\Psi_\sigma({\bf r})$ is a fermion field operator with
pseudo-spin $\sigma=\uparrow,\downarrow$, 
and $\mu$ is the chemical potential. $U$ is an
$s$-wave pairing interaction, and $V_{\rm trap}({\bf r})$ is an
anisotropic harmonic trap potential given by
\begin{equation}
V_{\rm trap}({\bf r})=\sum_{\alpha=x,y,z}{1 \over 2}m
\Omega_\alpha^2 r_\alpha^2.
\label{eq.2.2}
\end{equation}
The proof of Kohn's theorem explained in the introduction can be
easily extended to the second quantized Hamiltonian in 
Eq. (\ref{eq.2.1}). When we define
\begin{equation}
{\hat P}_\alpha\equiv \sum_\sigma\int d{\bf r}
\Psi_\sigma^\dagger({\bf r})
[m\Omega_\alpha {\hat r}_\alpha-i{\hat p}_\alpha]
\Psi_\sigma({\bf r})
~~~(\alpha=x,y,z),
\label{eq.2.2b}
\end{equation}
this operator satisfies $[H,{\hat P}_\alpha]=
\Omega_\alpha{\hat P}_\alpha$. Thus, the state
${\hat P}_\alpha|\Psi_0\rangle$ has 
the excitation energy $\Omega_\alpha$, if $|\Psi_0\rangle$ 
is the ground state wave-function.
\par
In the Hartree-Fock-Bogoliubov (HFB) approximation, Eq. (\ref{eq.2.1})
reduces to
\begin{eqnarray}
H_{\rm HFB}
&=&
\sum_\sigma\int d{\bf r}
\Psi^\dagger_\sigma({\bf r})
\Bigl[{\hat h}_0({\bf r})-{U \over 2}n({\bf r})
\Bigr]
\Psi_\sigma({\bf r})
+
\int d{\bf r}
[\Delta({\bf r})
\Psi_{\uparrow}^\dagger({\bf r})
\Psi_{\downarrow}^\dagger({\bf r})
+
{\rm h.c.}
],
\label{eq.2.3}
\end{eqnarray}
where ${\hat h}_0({\bf r})=-{\nabla^2 \over 2m}+V_{\rm trap}({\bf r})-\mu$ is
the Hamiltonian for non-interacting Fermi atoms. $H_{\rm HFB}$ 
includes two mean fields, involving the local density of Fermi
atoms $n({\bf r})\equiv
\sum_\sigma\langle\Psi_\sigma^\dagger({\bf r})\Psi_\sigma
({\bf r})\rangle$ as well as the off-diagonal field associated with
the Cooper-pair
order parameter $\Delta({\bf r})\equiv -U\langle
\Psi_\downarrow({\bf r})\Psi_\uparrow({\bf r})\rangle$\cite{noteN}. 
In the well-known Nambu representation, $H_{\rm HFB}$ is conveniently
written as\cite{Schrieffer}
\begin{eqnarray}
H_{\rm HFB}=
\int d{\bf r}
{\hat \Psi}^\dagger({\bf r})
\Bigr[
{\hat h}_0({\bf r})\sigma_3-{U \over 2}n({\bf r})\sigma_3+
\Delta({\bf r})\sigma_1
\Bigl]
{\hat \Psi}({\bf r}),
\label{eq.2.5}
\end{eqnarray}
where ${\hat \Psi}^\dagger({\bf r})\equiv(\Psi^\dagger_\uparrow({\bf r}),
\Psi_\downarrow({\bf r}))$ and
\begin{eqnarray}
{\hat \Psi}({\bf r})\equiv
\left(
\begin{array}{c}
\Psi_\uparrow({\bf r}) \\
\Psi^\dagger_\downarrow({\bf r})
\end{array}
\right)
\label{eq.2.5b}
\end{eqnarray}
are the two-component Nambu field operators, and $\sigma_\alpha$
($\alpha=1,2,3$) are the Pauli matrices. In Eq. (\ref{eq.2.5}), we take the
order parameter $\Delta({\bf r})$ real and proportional to the
$\sigma_1$-component. This choice is always possible, in the absence of
supercurrents or vortices.
\par
The HFB Hamiltonian in Eq. (\ref{eq.2.5}) can be diagonalized using
the solution of the well-known Bogoliubov de-Gennes (BdG) equations,
\begin{equation}
{\hat h}\Psi_n({\bf r})=E_n\Psi_n({\bf r}),
\label{eq.2.6}
\end{equation}
where 
\begin{equation}
{\hat h}\equiv {\hat h}_0({\bf r})\sigma_3-{U \over 2}
n({\bf r})\sigma_3+\Delta({\bf r})\sigma_1.
\label{eq.2.6b}
\end{equation}
The energy $E_n$ of the two-component wave-function
\begin{eqnarray}
\Psi_n({\bf r})=
\Bigl(
\begin{array}{c}
u_n({\bf r})\\
v_n({\bf r})
\end{array}
\Bigr)
\label{eq.2.7}
\end{eqnarray}
can be both positive and negative. Actually, a negative energy state
$\Psi_{n<0}$ ($-E_n<0$) is related to a positive energy state 
$\Psi_{n>0}$ ($E_n>0$) by
\begin{equation}
\Psi_{n<0}=i\sigma_2\Psi_{n>0}.
\label{eq.2.8}
\end{equation}
Thus, we need only solve for the positive energy solutions of the BdG
equations. 
\par
The Bogoliubov transformation\cite{noteBdG} is given by
\begin{equation}
{\hat \Psi}({\bf r})=\sum_{E_n>0}
[\Psi_n({\bf r})\gamma_{n\uparrow}
+i\sigma_2\Psi_n({\bf r})\gamma^\dagger_{n\downarrow}],
\label{eq.2.10b}
\end{equation}
in which case Eq. (\ref{eq.2.3}) can be diagonalized as
\begin{equation}
H_{\rm HFB}=\sum_{E_n>0,\sigma} 
E_n\gamma_{n\sigma}^\dagger\gamma_{n\sigma}. 
\label{eq.2.10c}
\end{equation}
Here $\gamma_{n\sigma}$ is the annihilation operator of a
Bogoliubov quasi-particle. When we define the fermion operators
$\gamma_{n>0}\equiv\gamma_{n\uparrow}~(E_n>0)$ and
$\gamma_{n<0}\equiv\gamma_{n\downarrow}^\dagger~(E_n<0)$,
Eqs. (\ref{eq.2.10b}) and
(\ref{eq.2.10c}) are rewritten, respectively, as
\begin{equation}
{\hat \Psi}({\bf r})=\sum_n\Psi_n({\bf r}){\hat \gamma}_n,
\label{eq.2.9}
\end{equation}
\begin{equation}
H_{\rm HFB}=\sum_n E_n{\hat \gamma}_n^\dagger{\hat \gamma}_n,
\label{eq.2.10}
\end{equation}
where the summations in Eqs. (\ref{eq.2.9}) and (\ref{eq.2.10}) are
 taken over all the eigenstates with both positive and negative
 energies.
In the following sections, we use the Bogoliubov transformation
given by Eq. (\ref{eq.2.9}).
\par
The two mean-fields $\Delta({\bf r})$ and 
$n({\bf r})$ are determined self-consistently by
\begin{eqnarray}
\Delta({\bf r})=
{U \over 2}\sum_{E_n>0}^{\omega_c}
\Psi^\dagger_n({\bf r})\sigma_1\Psi_n({\bf r})
[1-2f_n],
\label{eq.2.11}
\end{eqnarray}
\begin{equation}
n({\bf r})=-\sum_{E_n>0}\Psi^\dagger_n({\bf r})\sigma_3\Psi_n({\bf r})
[1-2f_n]+\delta(0).
\label{eq.2.12}
\end{equation}
Here $f_n$ is the Fermi distribution function with energy $E_n$. As
usual, we need a cutoff $\omega_c$ in the energy-summation of the gap
equation (\ref{eq.2.11}). The divergent term $\delta(0)$
in Eq. (\ref{eq.2.12}) is ultimately canceled out by 
the divergent term involved in the first term.
\vskip2mm
\subsection{Generalized random phase approximation (GRPA)}
\vskip2mm
\par
The static HFB Hamiltonian in Eq. (\ref{eq.2.3}) neglects fluctuations 
in both the density
and Cooper channels. 
These fluctuation effects can be included by 
including interactions left out in the HFB approximation.
To describe such interactions, it is convenient to introduce
a {\it generalized} density operator ${\hat \rho}_\alpha\equiv
{\hat \Psi}^\dagger({\bf r})\sigma_\alpha{\hat \Psi}({\bf r})$, where 
$\alpha=1,2,3$
represent the amplitude fluctuations of the order parameter, 
the phase fluctuations of the order parameter, and
density fluctuations, respectively\cite{note3}. Then interactions 
involving fluctuations in Eq. (\ref{eq.2.1}) can be written
as\cite{Ohashi4,OhashiT,note2}
\begin{eqnarray}
U_\alpha^{\rm FL}
\equiv
-{U \over 4}\int d{\bf r}{\hat \rho}_\alpha
({\bf r}){\hat \rho}_\alpha({\bf r})~~~(\alpha=1,2,3).
\label{eq.2.13}
\end{eqnarray}
Here $U_1^{\rm FL}$ and $U_2^{\rm FL}$ are interactions in the Cooper channel, while 
the interaction in the density channel is given by $U_3^{\rm FL}$.
\par
In the superfluid phase, density fluctuations couple with superfluid
fluctuations through the Josephson effect. The {\it generalized} random
phase approximation (GRPA) is a RPA kind of approximation which includes
this additional coupling\cite{OhashiT}. When we use the GRPA to treat 
$U_j^{\rm FL}$ in
Eq. (\ref{eq.2.13}) in a linear response calculation, we find that
the response in the generalized density is described by
\begin{eqnarray}
\delta\rho_\alpha({\bf r},\omega)=-{U \over 2}\sum_{\beta=1}^3
\int d{\bf r}'\Pi^0_{\alpha\beta}
({\bf r},{\bf r}',\omega)\delta\rho_\beta({\bf r}',\omega)~~~(\alpha=1,2,3).
\label{eq.2.14}
\end{eqnarray}
Here $\Pi^0_{\alpha\beta}$ is the zero-th order {\it generalized} density
correlation function defined by\cite{OhashiT}
\begin{equation}
\Pi^0_{\alpha\beta}({\bf r},{\bf r}',\omega)=
-i\int_0^\infty dt e^{i\omega t}
\langle
[{\hat \rho}_\alpha({\bf r},t),{\hat \rho}_\beta({\bf r}',0)]
\rangle.
\label{eq.2.15}
\end{equation}
As we discuss elsewhere\cite{Ohashi4}, the ordinary density 
correlation function is $\Pi^0_{33}$, while
$\Pi^0_{11}$ and $\Pi^0_{22}$ describe amplitude and phase
fluctuations of the 
order parameter, respectively. The off-diagonal components
of Eq. (\ref{eq.2.15}) represent coupling between these fluctuations. For
example, $\Pi^0_{23}$ expresses a phase-density coupling, originating
from the Josephson effect. Thus, Eq. (\ref{eq.2.14}) describes
a collective mode
in terms of the amplitude ($\delta\rho_1$) and phase
($\delta\rho_2$) oscillations in the Cooper-channel, 
as well as the ordinary density
oscillation ($\delta\rho_3$).
\par
In a uniform gas, where we can use $\delta\rho_\alpha({\bf r},\omega)
=e^{i{\bf q}\cdot{\bf r}}\delta\rho_\alpha({\bf q},\omega)$, 
Eq. (\ref{eq.2.14})
reduces to the $3\times 3$ matrix equation
\begin{eqnarray}
\Bigl[
1-{U \over 2}{\hat \Pi}^0({\bf q},\omega)
\Bigr]
\left(
\begin{array}{c}
\delta\rho_1({\bf q},\omega)\\
\delta\rho_2({\bf q},\omega)\\
\delta\rho_3({\bf q},\omega)\\
\end{array}
\right)
=0.
\label{eq.2.16}
\end{eqnarray}
Here ${\hat \Pi}^0({\bf q},\omega)\equiv\{\Pi^0_{\alpha\beta}
({\bf q},\omega)\}$ $(\alpha,\beta=1,2,3)$ 
is the $3\times3$-matrix correlation
function in momentum space (The detailed expressions for
$\Pi^0_{\alpha\beta}
({\bf q},\omega)$ are given in Ref. \cite{Ohashi4}.). The solution
of the $3\times 3$ matrix equation in Eq. (\ref{eq.2.16}) can be 
shown to be the
same as the poles of the ($3\times 3$ matrix) GRPA
density correlation function, ${\hat \Pi}({\bf q},\omega)
\equiv[1-{U \over 2}{\hat \Pi}^0({\bf q},\omega)]^{-1}{\hat \Pi}^0
({\bf q},\omega)$\cite{Ohashi4,Bruun1}. 
\par
In a harmonic trap, 
$\Pi_{\alpha\beta}^0({\bf r},{\bf r}',\omega)$ can be calculated
from the analytic continuation of the corresponding two-particle 
{\it thermal} Green's functions,
\begin{eqnarray}
{\tilde \Pi}^0_{\alpha\beta}({\bf r},{\bf r}',i\nu_n)
=
{1 \over \beta}\sum_{\omega_l}
{\rm Tr}
\Bigl[\sigma_\alpha{\hat G}({\bf r},{\bf r}',i\omega_l+i\nu_n)\sigma_\beta
{\hat G}({\bf r}',{\bf r},i\omega_l)
\Bigr],
\nonumber
\\
\label{eq.2.17}
\end{eqnarray}
where $i\omega_l$ and $i\nu_n$ are the usual fermion and boson Matsubara
frequencies, respectively. ${\hat G}({\bf r},{\bf r}',i\omega_l)$ is
the $2\times 2$-matrix single-particle 
thermal Green's function, given by
\begin{eqnarray}
{\hat G}({\bf r},{\bf r}',i\omega_l)
=
\sum_n{\Psi_n({\bf r})\Psi_n^\dagger({\bf r}') \over i\omega_l-E_n},
\label{eq.2.18}
\end{eqnarray}
where $\Psi_n({\bf r})$ are the Nambu eigenstate spinors defined in
Eq. (\ref{eq.2.6}).
After doing the $i\omega_l$-summation in Eq. (\ref{eq.2.17}), we
make the usual analytic continuation
$i\nu_n\to\omega_+\equiv\omega+i\delta$. 
The resulting diagonal correlation function 
is given by
\begin{eqnarray}
\Pi^0_{\alpha\alpha}({\bf r},{\bf r}',\omega)
&=&
\sum_{nn'}
{E_{n'}-E_n \over \omega_+^2-(E_{n'}-E_n)^2}
[f_n-f_{n'}]
\Bigl[
\Psi^\dagger_{n'}({\bf r})\sigma_\alpha\Psi_n({\bf r})
\Bigr]
\Bigl[
\Psi^\dagger_n({\bf r}')\sigma_\alpha\Psi_{n'}({\bf r}')
\Bigr]
,
\nonumber
\\
\label{eq.2.19}
\end{eqnarray}
where we have used the relation between positive and
negative energy states in Eq. (\ref{eq.2.8}). Similarly, 
the off-diagonal correlation
functions $\Pi^0_{\alpha\ne \beta}$ involving coupling between
fluctuations are given by
\begin{eqnarray}
\Pi^0_{12}({\bf r},{\bf r}',\omega)
&=&
-\sum_{nn'}
{\omega_+ \over \omega_+^2-(E_{n'}-E_n)^2}
[f_n-f_{n'}]
\Bigl[
\Psi^\dagger_{n'}({\bf r})\sigma_1\Psi_n({\bf r})
\Bigr]
\Bigl[
\Psi^\dagger_n({\bf r}')\sigma_2\Psi_{n'}({\bf r}')
\Bigr],
\nonumber
\\
\label{eq.2.20}
\end{eqnarray}
\begin{eqnarray}
\Pi^0_{21}({\bf r},{\bf r}',\omega)
&=&
-\sum_{nn'}
{\omega_+ \over \omega_+^2-(E_{n'}-E_n)^2}
[f_n-f_{n'}]
\Bigl[
\Psi^\dagger_{n'}({\bf r})\sigma_2\Psi_n({\bf r})
\Bigr]
\Bigl[
\Psi^\dagger_n({\bf r}')\sigma_1\Psi_{n'}({\bf r}')
\Bigr],
\nonumber
\\
\label{eq.2.21}
\end{eqnarray}
\begin{eqnarray}
\Pi^0_{23}({\bf r},{\bf r}',\omega)
&=&
-\sum_{nn'}
{\omega_+ \over \omega_+^2-(E_{n'}-E_n)^2}
[f_n-f_{n'}]
\Bigl[
\Psi^\dagger_{n'}({\bf r})\sigma_2\Psi_n({\bf r})
\Bigr]
\Bigl[
\Psi^\dagger_n({\bf r}')\sigma_3\Psi_{n'}({\bf r}')
\Bigr],
\nonumber
\\
\label{eq.2.22}
\end{eqnarray}
\begin{eqnarray}
\Pi^0_{32}({\bf r},{\bf r}',\omega)
&=&
-\sum_{nn'}
{\omega_+ \over \omega_+^2-(E_{n'}-E_n)^2}
[f_n-f_{n'}]
\Bigl[
\Psi^\dagger_{n'}({\bf r})\sigma_3\Psi_n({\bf r})
\Bigr]
\Bigl[
\Psi^\dagger_n({\bf r}')\sigma_2\Psi_{n'}({\bf r}')
\Bigr],
\nonumber
\\
\label{eq.2.23}
\end{eqnarray}
\begin{eqnarray}
\Pi^0_{13}({\bf r},{\bf r}',\omega)
&=&
\sum_{nn'}
{E_{n'}-E_n \over \omega_+^2-(E_{n'}-E_n)^2}
[f_n-f_{n'}]
\Bigl[
\Psi^\dagger_{n'}({\bf r})\sigma_1\Psi_n({\bf r})
\Bigr]
\Bigl[
\Psi^\dagger_n({\bf r}')\sigma_3\Psi_{n'}({\bf r}')
\Bigr],
\nonumber
\\
\label{eq.2.24}
\end{eqnarray}
\begin{eqnarray}
\Pi^0_{31}({\bf r},{\bf r}',\omega)
&=&
\sum_{nn'}
{E_{n'}-E_n \over \omega_+^2-(E_{n'}-E_n)^2}
[f_n-f_{n'}]
\Bigl[
\Psi^\dagger_{n'}({\bf r})\sigma_3\Psi_n({\bf r})
\Bigr]
\Bigl[
\Psi^\dagger_n({\bf r}')\sigma_1\Psi_{n'}({\bf r}')
\Bigr].
\nonumber
\\
\label{eq.2.25}
\end{eqnarray}
\par
When we substitute Eqs. (\ref{eq.2.19})-(\ref{eq.2.25}) into
Eq. (\ref{eq.2.14}), we obtain the following coupled equations for the
collective modes
\begin{eqnarray}
\delta\rho_1({\bf r},\omega)
&=&
-{U \over 2}\sum_{nn'}
{f_n-f_{n'} \over \omega_+^2-(E_{n'}-E_n)^2}
\Bigl[\Psi_{n'}^\dagger({\bf r})\sigma_1\Psi_n({\bf r})\Bigr]
\Bigl[
(E_{n'}-E_n)
\langle n|\sigma_1\delta\rho_1({\bf r}',\omega)
|n'\rangle
\nonumber
\\
&-&
\omega_+
\langle n|\sigma_2\delta\rho_2({\bf r}',\omega)
|n'\rangle
+
(E_{n'}-E_n)
\langle n|\sigma_3\delta\rho_3({\bf r}',\omega)
|n'\rangle
\Bigr],
\label{eq.2.26}
\end{eqnarray}
\begin{eqnarray}
\delta\rho_2({\bf r},\omega)
&=&
-{U \over 2}\sum_{nn'}
{f_n-f_{n'} \over \omega_+^2-(E_{n'}-E_n)^2}
\Bigl[\Psi_{n'}^\dagger({\bf r})\sigma_2\Psi_n({\bf r})\Bigr]
\Bigl[
-\omega_+
\langle n|\sigma_1\delta\rho_1({\bf r}',\omega)
|n'\rangle
\nonumber
\\
&+&
(E_{n'}-E_n)
\langle n|\sigma_2\delta\rho_2({\bf r}',\omega)
|n'\rangle
-
\omega_+
\langle n|\sigma_3\delta\rho_3({\bf r}',\omega)
|n'\rangle
\Bigr],
\label{eq.2.27}
\end{eqnarray}
\begin{eqnarray}
\delta\rho_3({\bf r},\omega)
&=&
-{U \over 2}\sum_{nn'}
{f_n-f_{n'} \over \omega_+^2-(E_{n'}-E_n)^2}
\Bigl[\Psi_{n'}^\dagger({\bf r})\sigma_3\Psi_n({\bf r})\Bigr]
\Bigl[
(E_{n'}-E_n)
\langle n|\sigma_1\delta\rho_1({\bf r}',\omega)
|n'\rangle
\nonumber
\\
&-&
\omega_+
\langle n|\sigma_2\delta\rho_2({\bf r}',\omega)
|n'\rangle
+
(E_{n'}-E_n)
\langle n|\sigma_3\delta\rho_3({\bf r}',\omega)
|n'\rangle
\Bigr].
\label{eq.2.28}
\end{eqnarray}
Here we have introduced the notation
\begin{equation}
\langle n|\sigma_\alpha\delta\rho_\alpha({\bf r}')|n'\rangle\equiv
\int d{\bf r}'\Psi^\dagger_n({\bf r}')
\sigma_\alpha\delta\rho_\alpha({\bf r}')\Psi_{n'}({\bf r}').
\label{eq.2.29}
\end{equation}
In HFB-GRPA formalism, the collective modes are determined by
the solutions of Eqs. (\ref{eq.2.26})-(\ref{eq.2.28}). The single particle
excitations $E_n$, particle density $n({\bf r})$, and order parameter
$\Delta({\bf r})$ are obtained  by solving Eqs. (\ref{eq.2.6}),
(\ref{eq.2.11}) and (\ref{eq.2.12}) self-consistently. 
\vskip2mm
\subsection{Goldstone mode}
\par
Since we have assumed a short-range effective pairing interaction in
Eq. (\ref{eq.2.1}), a cutoff $\omega_c$ is necessary in calculating
correlation functions related to superfluid fluctuations, e.g.,
$\Pi^0_{22}$. 
A similar cutoff is needed in solving the gap equation (\ref{eq.2.11}).
However, we must take some care in how we introduce this
same cutoff $\omega_{\rm c}$ in calculating the correlation functions.
For this purpose, a crucial role is played by involving 
Goldstone's theorem, which describes the appearance of an excitation 
in the ordered state associated with a broken continuous
symmetry. We want our HFB-GRPA formalism to satisfy 
this theorem and thus it is reasonable to introduce 
$\omega_c$ to ensure this happens.
\par
In the superfluid phase, the ground state is chosen from
 an infinitely large number of  
degenerate candidates for ground state, 
that are characterized by the phase of 
order parameter $\Delta({\bf r})e^{i\phi}$ ($0\le \phi<2\pi$).
The Goldstone mode is the excitation from an assumed ground state
 (where we set $\phi=0$) to the 
other degenerate states ($\phi\ne 0$)\cite{Anderson}. 
As a result, the Goldstone mode is physically described 
by a phase oscillation 
of the order parameter with a zero excitation energy.
The phase fluctuations of order parameter described by
$\delta\rho_2({\bf r},\omega=0)$ in 
Eq. (\ref{eq.2.27}) are indeed decoupled
from other fluctuations at $\omega=0$, because
one can show $\Pi^0_{12}=\Pi^0_{21}=\Pi^0_{23}=\Pi^0_{32}=0$ (see
Eqs. (\ref{eq.2.20})-(\ref{eq.2.23})). Taking
$(\delta\rho_1,\delta\rho_2,\delta\rho_3)=(0,\delta\rho_2,0)$,
we are left with a single equation at $\omega=0$,
\begin{eqnarray}
\delta\rho_2({\bf r},0)
&=&
-{U \over 2}
\sum_{nn'}
{f_{n'}-f_n \over E_{n'}-E_n}
\Bigl[
\Psi_{n'}^\dagger({\bf r})\sigma_2\Psi_n({\bf r})
\Bigr]
\langle n|\sigma_2\delta\rho_2({\bf r}',0)|n'\rangle.
\label{eq.2.30}
\end{eqnarray}
\par
We introduce a cutoff $\omega_c$ in the energy-summation
($n$-summation) by attaching the step function
$\Theta(\omega_c-|E_n|)$ to the Fermi distribution functions $f_n$ and
$f_{n'}$ in Eq. (\ref{eq.2.30}) as follows;
\begin{eqnarray}
f_n&\to&{\tilde f}_n\equiv \Theta(\omega_c-|E_n|)f_n,
\nonumber
\\
f_{n'}&\to&{\tilde f}_{n'}\equiv \Theta(\omega_c-|E_{n'}|)f_{n'}.
\label{eq.2.31}
\end{eqnarray}
To prove that this prescription is consistent with the BCS gap equation
(\ref{eq.2.11}), we next show that $\delta\rho_2({\bf r},0)=\Delta({\bf r})$
is a solution of Eq. (\ref{eq.2.30}) {\it when 
the order parameter satisfies Eq. (\ref{eq.2.11})}. From the commutation
relation $[{\hat h},\sigma_3]=-2i\Delta({\bf r})\sigma_2$ (where
${\hat h}$ is defined in Eq. (\ref{eq.2.6b})), we
obtain 
\begin{eqnarray}
\langle n|\sigma_2\Delta({\bf r}')|n'\rangle
=
{i \over 2}\langle n|[{\hat h},\sigma_3]|n'\rangle
={i \over 2}(E_n-E_{n'})\langle n|\sigma_3|n'\rangle.
\label{eq.2.32}
\end{eqnarray}
Substituting  $\delta\rho_2({\bf r},0)=\Delta({\bf r})$ into
the RHS of Eq. (\ref{eq.2.30}) ($\equiv S({\bf r},0))$, we find
\begin{eqnarray}
S({\bf r},0)
&=&
{iU \over 4}
\sum_{nn'}
[{\tilde f}_{n'}-{\tilde f}_n]
\Bigl[
\Psi_{n'}^\dagger({\bf r})\sigma_2\Psi_n({\bf r})
\Bigr]
\langle n|\sigma_3|n'\rangle
\nonumber
\\
&=&
{iU \over 4}
\sum_n
{\tilde f}_n\Psi_{n}^\dagger({\bf r})[\sigma_2,\sigma_3]\Psi_n({\bf r})
\nonumber
\\
&=&
-{U \over 2}
\sum_n
{\tilde f}_n\Psi_{n}^\dagger({\bf r})\sigma_1\Psi_n({\bf r})
\nonumber
\\
&=&
-{U \over 2}
\sum_{E_n>0}^{\omega_c}
\Bigl[
\Psi_{n}^\dagger({\bf r})\sigma_1\Psi_n({\bf r})f_n
+
\Psi_{n}^\dagger({\bf r})\sigma_2\sigma_1\sigma_2\Psi_n({\bf r})(1-f_n)
\Bigr]
\nonumber
\\
&=&
\sum_{E_n>0}^{\omega_c}
\Psi_{n}^\dagger({\bf r})\sigma_1\Psi_n({\bf r})[1-2f_n].
\label{eq.2.33}
\end{eqnarray}
In this calculation, we have used the completeness condition,
$\sum_n\Psi_n({\bf r})\Psi_n^\dagger({\bf r}')
=\delta({\bf r}'-{\bf r})$. From the gap equation (\ref{eq.2.11}),
Eq. (\ref{eq.2.33}) 
is found to be precisely equal to $\Delta({\bf r})$. 
\par
As a result, we have shown that at $\omega=0$,
\begin{eqnarray}
\left(
\begin{array}{c}
\delta\rho_1({\bf r},0)\\
\delta\rho_2({\bf r},0)\\
\delta\rho_3({\bf r},0)
\end{array}
\right)
=
\left(
\begin{array}{c}
0 \\
\Delta({\bf r})\\
0
\end{array}
\right),
\label{eq.2.33b}
\end{eqnarray}
which describes the Goldstone mode. We conclude that 
Eq. (\ref{eq.2.31}) is 
consistent choice of the cutoff $\omega_c$
with the gap equation (\ref{eq.2.11}), in the sense that the
HFB-GRPA leads to linear response functions which satisfy Goldstone's 
theorem.  In the following sections, we use
Eq. (\ref{eq.2.31}) in evaluating the superfluid fluctuations
described by $\delta\rho_1$ and $\delta\rho_2$ in Eqs. (\ref{eq.2.26})
and (\ref{eq.2.27}). 
\vskip2mm
\subsection{Kohn mode in Trapped Fermi superfluids}
\vskip2mm
When the frequency of a collective mode is non-zero, density
fluctuations couple with superfluid fluctuations, so that we have to
solve the {\it coupled} equations (\ref{eq.2.26})-(\ref{eq.2.28}). In
this section, we prove that
\begin{eqnarray}
\left(
\begin{array}{c}
\delta\rho_1({\bf r},\Omega_x) \\
\delta\rho_2({\bf r},\Omega_x) \\
\delta\rho_3({\bf r},\Omega_x)
\end{array}
\right)
=
\left(
\begin{array}{c}
\partial_x\Delta({\bf r}) \\
-2im\Omega_x x\Delta({\bf r}) \\
-{U \over 2}\partial_xn({\bf r})
\end{array}
\right)
\label{eq.2.34}
\end{eqnarray}
are the solutions of Eqs. (\ref{eq.2.26})-(\ref{eq.2.28}), with frequency
$\omega$ equal to the trap frequency $\Omega_x$. We discuss the 
physical meaning
of these solutions after giving the proof.
\par
From the
two commutation relations $[x,{\hat h}]={1 \over m}\sigma_3\partial_x$
and $[\partial_x,
{\hat h}]=\sigma_3m\Omega_x^2x+\sigma_1\partial_x\Delta({\bf r})
-{U \over 2}\sigma_3\partial_xn({\bf r})$, we obtain
\begin{eqnarray}
(E_{n'}-E_n)
\langle n|\sigma_3x|n'\rangle
=
2i
\langle n|\sigma_2\Delta({\bf r})x|n'\rangle
+
{1 \over m}
\langle n|\partial_x|n'\rangle,
\label{eq.2.35}
\end{eqnarray}
\begin{eqnarray}
(E_{n'}-E_n)
\langle n|\partial_x|n'\rangle
=
m\Omega_x^2\langle n|\sigma_3x|n'\rangle
+
\langle n|\sigma_1(\partial_x\Delta({\bf r}))|n'\rangle
-
{U \over 2}
\langle n|\sigma_3(\partial_xn({\bf r}))|n'\rangle.
\label{eq.2.36}
\end{eqnarray}
These two equations give
\begin{eqnarray}
\Bigl[
(E_{n'}-E_n)^2-\Omega_x^2
\Bigr]
\langle n|\partial_x|n'\rangle
&=&
(E_{n'}-E_n)\langle n|\sigma_1(\partial_x\Delta({\bf r}))|n\rangle
\nonumber
\\
&+&
2im\Omega_x^2\langle n|\sigma_2\Delta({\bf r})x|n\rangle
\nonumber
\\
&-&
(E_{n'}-E_n)\langle n|\sigma_3{U \over 2}(\partial_xn({\bf r}))|n\rangle,
\nonumber
\\
\label{eq.2.37}
\end{eqnarray}
and
\begin{eqnarray}
m\Bigl[
(E_{n'}-E_n)^2-\Omega_x^2
\Bigr]
\langle n|\sigma_3x|n'\rangle
&=&
\langle n|\sigma_1\Delta({\bf r})x|n\rangle
\nonumber
\\
&+&
2im(E_{n'}-E_n)
\langle n|\sigma_2\Delta({\bf r})x|n\rangle
\nonumber
\\
&-&
\langle n|\sigma_3{U \over 2}(\partial_xn({\bf r}))|n\rangle.
\nonumber
\\
\label{eq.2.38}
\end{eqnarray}
We next 
substitute Eq. (\ref{eq.2.34}) into the RHS of (\ref{eq.2.26}), which we
denote as $S_1(\omega)$. At $\omega=\Omega_x$, $S_1$ can be reduced to, by
using Eq. (\ref{eq.2.37}),
\begin{eqnarray}
S_1(\Omega_x)
&=&
{U \over 2}\sum_{nn'}[{\tilde f}_n-{\tilde f}_{n'}]
\Bigl[
\Psi_{n'}^\dagger({\bf r})\sigma_1\Psi_n({\bf r})
\Bigr]
\langle n|\partial_{x'}|n'\rangle
\nonumber
\\
&=&
-{U \over 2}\sum_n{\tilde f}_n
\Bigl[
(\partial_x\Psi_n^\dagger({\bf r}))\sigma_1\Psi_n({\bf r})
+
\Psi_n^\dagger({\bf r})\sigma_1\partial_x\Psi_n({\bf r})
\Bigr]
\nonumber
\\
&=&
-\partial_x{U \over 2}\sum_n{\tilde f}_n\Psi_n^\dagger
({\bf r})\sigma_1\Psi_n({\bf r})
\nonumber
\\
&=&
-\partial_x{U \over 2}\sum_{E_n>0}^{\omega_c}
\Bigl[
\Psi_n^\dagger({\bf r})\sigma_1\Psi_n({\bf r})f_n
+
\Psi_n^\dagger({\bf r})\sigma_2\sigma_1\sigma_2\Psi_n({\bf r})(1-f_n)
\Bigr]
\nonumber
\\
&=&
\partial_x{U \over 2}\sum_{E_n>0}^{\omega_c}
\Psi_n^\dagger({\bf r})\sigma_1\Psi_n({\bf r})[1-2f_n]
\nonumber
\\
&=&
\partial_x\Delta({\bf r})~~~~(=\delta\rho_1({\bf r},\Omega_x)).
\label{eq.2.39}
\end{eqnarray}
Thus our trial solution in Eq. (\ref{eq.2.34}) satisfies 
Eq. (\ref{eq.2.26}) for
$\omega=\Omega_x$. 
\par
In the same way, we also find that
Eq. (\ref{eq.2.34}) is a solution of Eq. (\ref{eq.2.28}). Indeed,
substituting Eq. (\ref{eq.2.37}) into the RHS of Eq. (\ref{eq.2.28})
with $\omega=\Omega_x$ ($\equiv S_3(\Omega_x)$), we obtain 
\begin{eqnarray}
S_3(\Omega_x)
&=&
{U \over 2}\sum_{nn'}[f_n-f_{n'}]
\Bigl[
\Psi_{n'}^\dagger({\bf r})\sigma_3\Psi_n({\bf r})
\Bigr]
\langle n|\partial_{x'}|n'\rangle
\nonumber
\\
&=&
-{U \over 2}\partial_x\sum_nf_n
\Psi_n^\dagger({\bf r})\sigma_3\Psi_n({\bf r})
\nonumber
\\
&=&
{U \over 2}\partial_x\sum_{E_n>0}
\Psi_n^\dagger({\bf r})\sigma_3\Psi_n({\bf r})(1-2f_n)
\nonumber
\\
&=&
-{U \over 2}\partial_xn({\bf r})~~~~(\equiv\delta\rho_3({\bf r},\Omega_x)).
\label{eq.2.40}
\end{eqnarray}
The RHS of Eq. (\ref{eq.2.27}) ($\equiv S_2(\Omega_x)$) can be
evaluated by using Eq. (\ref{eq.2.38}). We find, for $\omega=\Omega_x$,
\begin{eqnarray}
S_2(\Omega_x)
&=&
-{U \over 2}m\Omega_x\sum_{nn'}[{\tilde f}_n-{\tilde f}_{n'}]
\Bigl[
\Psi_{n'}^\dagger({\bf r})\sigma_2\Psi_n({\bf r})
\Bigr]
\langle n|\sigma_3x|n'\rangle
\nonumber
\\
&=&
iUm\Omega_x\sum_n{\tilde f}_n\Psi_n({\bf r})\sigma_1\Psi_n({\bf r})x
\nonumber
\\
&=&
-iUm\Omega_x\sum_{E_n>0}^{\omega_c}\Psi_n({\bf r})\sigma_1\Psi_n({\bf r})(1-2f_n)
\nonumber
\\
&=&
-2im\Omega_x\Delta({\bf r})x~~~~(\equiv \delta\rho_2({\bf r},\Omega_x)).
\label{eq.2.41}
\end{eqnarray}
\par
From the results in Eqs. (\ref{eq.2.39})-(\ref{eq.2.41}), we 
have shown explicitly that fluctuations as given in 
Eq. (\ref{eq.2.34}) are indeed a solution of the coupled equations
(\ref{eq.2.26})-(\ref{eq.2.28}) with frequency $\Omega_x$. This is
the Kohn mode. Our calculation shows that in the BCS
pairing approximation, the Kohn mode at the trap frequency 
is described as a coupled
oscillation of density, phase, and amplitude of order parameter. 
We note that this contrasts with the Goldstone mode 
at $\omega=0$, which is associated 
with a pure phase
oscillation. 
\par
In the study of BEC of trapped Bose gases, 
it has been shown that the Kohn mode is
the rigid center of mass oscillation of the static condensate and 
non-condensate distributions\cite{Griffin2}. 
We now discuss the physics of the Kohn mode in Fermi superfluids, as
described by Eq. (\ref{eq.2.34}).
In fermion superfluidity,
the total local density profile at $t$ is given by
\begin{equation}
n({\bf r},t)=n({\bf r})+C 
Re[\delta\rho_3({\bf r},\Omega_x)e^{i\Omega_x t}],
\label{eq.2.41b}
\end{equation}
where $C$ is a constant determining the amplitude of the oscillation. 
Substitute the third component of Eq. (\ref{eq.2.34}) 
into Eq. (\ref{eq.2.41b}), we obtain
\begin{eqnarray}
n({\bf r},t)
&=&
n({\bf r})-C{U \over 2}\partial_x n({\bf r})\cos(\Omega_x t)
\nonumber
\\
&\simeq&
n({\bf r}-{\bf e}_xC{U \over 2}\cos(\Omega_x t)).
\label{eq.2.41b2}
\end{eqnarray}
Here ${\bf e}_x$ is the unit vector in the $x$-direction.
The Kohn mode solution in Fermi superfluids is thus found 
to be the center of mass motion of the total
density, just in the case in Bose gases\cite{Griffin2}. 
Similarly, the oscillation
of the order parameter is given by
\begin{equation}
\Delta({\bf r},t)=\Delta({\bf r})+C\delta\Delta({\bf r},t).
\label{eq.2.41c}
\end{equation}
The second term is related the amplitude ($\delta\rho_1$) and 
phase oscillations ($\delta\rho_2$) of the
order parameter, given by
\begin{equation}
\delta\Delta({\bf r},t)=
-{U \over 2}
\Bigl[
Re[\delta\rho_1({\bf r},\Omega_x)e^{i\Omega_x t}]
-i
Re[\delta\rho_2({\bf r},\Omega_x)e^{i\Omega_x t}]
\Bigr].
\label{eq.2.41d}
\end{equation}
Using the first and second components in Eq. (\ref{eq.2.34}), we find
(working to first order in the fluctuations)
\begin{eqnarray}
\Delta({\bf r},t)
&=&
\Delta({\bf r})-C{U \over 2}\partial_x\Delta({\bf r})\cos(\Omega_x t)
+iCUmx\Omega_x\Delta({\bf r})\sin(\Omega_x t)
\nonumber
\\
&\simeq&
\Delta({\bf r}-{\bf e}_xC{U \over 2}\cos(\Omega_x t))
e^{iCUmx\Omega_x\sin(\Omega_x t)}.
\label{eq.2.41e}
\end{eqnarray}
Since $|\Delta({\bf r}, t)|^2$ describes the density of 
the Cooper-pair condensate, 
we find that
this condensate oscillates in the same way as the total density profile
given by Eq. (\ref{eq.2.41b2}). Equation (\ref{eq.2.41e}) also shows that
the Kohn mode is accompanied by a phase oscillation with the
same frequency $\Omega_x$. This is due to the Josephson
effect, which couples the density oscillation with the phase 
oscillation of the Cooper-pair
order parameter.
\par
It is easy to show that the other two Kohn modes
with the trap frequencies $\Omega_y$ and $\Omega_z$ are, respectively, given by
\begin{eqnarray}
\left(
\begin{array}{c}
\delta\rho_1({\bf r},\Omega_y) \\
\delta\rho_2({\bf r},\Omega_y) \\
\delta\rho_3({\bf r},\Omega_y)
\end{array}
\right)
=
\left(
\begin{array}{c}
\partial_y\Delta({\bf r}) \\
-2im\Omega_y y\Delta({\bf r}) \\
-{U \over 2}\partial_yn({\bf r})
\end{array}
\right),
\label{eq.2.43}
\end{eqnarray}
\begin{eqnarray}
\left(
\begin{array}{c}
\delta\rho_1({\bf r},\Omega_z) \\
\delta\rho_2({\bf r},\Omega_z) \\
\delta\rho_3({\bf r},\Omega_z)
\end{array}
\right)
=
\left(
\begin{array}{c}
\partial_z\Delta({\bf r}) \\
-2im\Omega_z z\Delta({\bf r}) \\
-{U \over 2}\partial_zn({\bf r})
\end{array}
\right).
\label{eq.2.44}
\end{eqnarray}
\par
The above proof can be immediately extended to
the strong-coupling theory developed by Nozi\`eres and
Schmitt-Rink\cite{Nozieres} at $T_{\rm c}$ 
in the context of superconductivity. In
the NSR theory, fluctuations around the mean-field order parameter are
taken into account within the Gaussian
approximation\cite{Engelbrecht}. One solves the gap equation
together with the equation for the number of particles, which includes
the effect of fluctuations around the mean field approximation. 
While the gap equation has the same form as
Eq. (\ref{eq.2.11}) in the NSR theory, the chemical potential $\mu$
can be very different from the Fermi energy in the strong-coupling
regime. When we consider collective modes in NSR-GRPA 
theory, we again obtain the 
linear response equations in Eqs. (\ref{eq.2.26})-
(\ref{eq.2.28}). The only difference is that the chemical potential 
is now determined by the equation for the number of particles, including the
fluctuation effect in the Cooper channel. However, 
we note that the above proof based on 
Eqs. (\ref{eq.2.26})-(\ref{eq.2.28}) is always valid irrespective of the
value of the chemical potential. Thus, even if the chemical
potential remarkably deviates from the Fermi energy due to the
strong-coupling effect in the NSR-GRPA theory, we again
obtain the Kohn modes with the frequencies $\Omega_\alpha$ 
($\alpha=x,y,z$), as well as the zero frequency Goldstone mode.
Thus NSR-GRPA linear response formalism 
is found to be consistent with both Kohn's theorem 
and Goldstone's theorem, and can be used to 
study the BCS-BEC crossover, without violating these 
important theorems. This is a crucial requirement of any approximate 
many-body calculation of response functions.
\vskip2mm
\par
\section{Kohn mode in the presence of a Feshbach resonance}
\par
\vskip2mm
\subsection{Coupled fermion-boson model}
\vskip2mm
A Feshbach resonance can be used to produce a
strong attractive (pairing) interaction in trapped Fermi
gases\cite{Jin1,Ketterle1,Ohara,Jin4}.
The Feshbach resonance is associated with a dimer bound state, 
which is a boson.
This boson can enhance the pairing interaction between atoms. To include
this effect in a simple way, the coupled fermion-boson
model\cite{Timmermans1,Timmermans2,Holland1,Holland2,Kokkelmans1,
Ohashi1,Ohashi3,Ohashi4,Milstein,Lee,Ranninger}
is very convenient, which is given by
\begin{eqnarray}
H
&=&
\sum_\sigma\int d{\bf r}
\Psi^\dagger_\sigma({\bf r})[-{\nabla^2 \over 2m}+V_{\rm trap}-\mu]
\Psi_\sigma({\bf r})
+
\int d{\bf r}
\Phi^\dagger({\bf r})
[-{\nabla^2 \over 2M}+2\nu+V^B_{\rm trap}-\mu_B]\Phi({\bf r})
\nonumber
\\
&+&
g_{\rm r}\int d{\bf r}
\Bigl[
\Phi^\dagger({\bf r})
\Psi_{\downarrow}({\bf r})
\Psi_{\uparrow}({\bf r})+h.c.
\Bigr]
-
U\int d{\bf r}
\Psi_{\uparrow}^\dagger({\bf r})
\Psi_{\downarrow}^\dagger({\bf r})
\Psi_{\downarrow}({\bf r})
\Psi_{\uparrow}({\bf r}).
\label{eq.3.1}
\end{eqnarray}
Here $\Phi({\bf r})$ is a molecular Bose field operator describing
molecules associated with the Feshbach resonance. $g_{\rm r}$ is
the coupling constant of the Feshbach resonance, in which two Fermi atoms
can form one Bose molecule. In turn, the molecule can dissociate 
into two Fermi
atoms. $2\nu$ is the threshold energy of the Feshbach resonance; the
value of $2\nu$ can be varied by an external magnetic field. In
this model, $U$ describes a non-resonant attractive interaction. Since one
molecule consists of two Fermi atoms, we take $M=2m$. We also impose
the conservation of the total number of atoms as 
$N=N_{\rm F}+2N_{\rm B}$, 
where $N_{\rm F}$ and $N_{\rm B}$ represent the number of Fermi
atoms and Bose molecules, respectively. This condition has been
included in Eq. (\ref{eq.3.1}), with the chemical potentials $\mu$ and
$\mu_B\equiv2\mu$. 
$V_{\rm trap}^B$ is a harmonic trap
potential for molecules, given by
\begin{equation}
V_{\rm trap}^B=\sum_{\alpha=x,y,z}{1 \over 2}M{\Omega^B_\alpha}^2r_\alpha^2.
\label{eq.3.2}
\end{equation}
If the molecules and atoms feel the same trap frequency, we have
$\Omega^B_\alpha=\Omega_\alpha$. 
When the hyperfine states of atoms involved in the
molecule are different from the atomic hyperfine states
$\Psi_\sigma({\bf r})$, the molecule may feel a different trap
frequency.
\par
Although the coupled fermion-boson model in Eq. (\ref{eq.3.1}) has the
different form from Eqs. (\ref{eq.1.1}) and (\ref{eq.2.1}) due to the
presence of molecular bosons, we can show that 
Kohn's theorem is exactly satisfied using Eq. (\ref{eq.3.1}) 
in the special case when
$\Omega_\alpha^B=\Omega_\alpha$. The generators to excite the Kohn modes
are now given by
\begin{eqnarray}
{\hat P}_\alpha
&\equiv&
 \sum_\sigma\int d{\bf r}
\Psi_\sigma^\dagger({\bf r})
[m\Omega_\alpha {\hat r}_\alpha-i{\hat p}_\alpha]
\Psi_\sigma({\bf r})
\nonumber
\\
&+&
\int d{\bf r}
\Phi^\dagger({\bf r})
[M\Omega^B_\alpha {\hat r}_\alpha-i{\hat p}_\alpha]
\Phi({\bf r}).
~~~(\alpha=x,y,z)
\label{eq.3.2b}
\end{eqnarray}
Using Eqs. (\ref{eq.3.1}) and (\ref{eq.3.2b}), one can derive 
$[H,{\hat P}_\alpha]=\Omega_\alpha{\hat P}_\alpha$ ($\alpha=x,y,z$).
Thus ${\hat P}_\alpha|\Psi_0\rangle$ ($\alpha=1,2,3$) describe 
excited states with frequency $\Omega_\alpha$, when $|\Psi_0\rangle$
is the ground state wavefunction.
\par
The HFB Hamiltonian for the coupled fermion-boson model in
Eq. (\ref{eq.3.1}) has the form
\begin{equation}
H_{\rm HFB}=H_{\rm HFB}^F+H_{\rm HFB}^B.
\label{eq.3.3}
\end{equation}
Here the fermion Hamiltonian $H_{\rm HFB}^F$ has the same form as the BCS
Hamiltonian in Eq. (\ref{eq.2.3}), except that the Cooper-pair order parameter
$\Delta({\bf r})$ is now replaced with the composite order parameter,
\cite{Timmermans2,Holland1,Ohashi1,Ohashi4} 
\begin{equation}
{\tilde \Delta}({\bf r})\equiv\Delta({\bf r})+g_{\rm r}\phi_m({\bf r}),
\label{eq.3.4}
\end{equation} 
consisting of the Cooper-pair $\Delta({\bf r})$ plus the molecular
condensate $\phi_m({\bf r})\equiv\langle\Phi({\bf r})\rangle$. In
the equilibrium state, these two order parameters are related to each
other by the identity\cite{Ohashi5}
\begin{equation}
{g_{\rm r} \over U}\Delta({\bf r})={\hat h}_B\phi_m({\bf r}),
\label{eq.3.5}
\end{equation}
where ${\hat h}_B\equiv-{\nabla^2 \over 2M}+2\nu
+V_{\rm trap}^B({\bf r})-2\mu$. The Bose Hamiltonian 
$H_{\rm HFB}^B$ is given by
\begin{equation}
H_{\rm HFB}^B=
\int d{\bf r}
\delta\Phi^\dagger({\bf r}){\hat h}_B\delta\Phi({\bf r}),
\nonumber
\label{eq.3.6}
\end{equation}
where $\delta\Phi({\bf r})\equiv\Phi({\bf r})-
\langle\Phi({\bf r})\rangle=\Phi({\bf r})-\phi_m$.
The identity in Eq. (\ref{eq.3.5}) plays a crucial role in the
following discussion. 
\par
The BdG equations have the same form as Eq. (\ref{eq.2.6}), except that now
$\Delta({\bf r})$ in Eq. (\ref{eq.2.6b}) is replaced with the composite order
parameter ${\tilde \Delta}({\bf r})$. Using the solution of the BdG equations,
the Cooper-pair order parameter $\Delta({\bf r})$ and the Fermi atom density
$n({\bf r})$ are calculated from Eqs. (\ref{eq.2.11}) and
(\ref{eq.2.12}), respectively. The equilibrium 
molecular condensate $\phi_m({\bf
  r})$ is then obtained from Eq. (\ref{eq.3.5}).
\par
The presence of 
molecular bosons leads to an effective interaction between Fermi
atoms. In GRPA, this interaction introduces an additional term to
the linear response equation (see Eq. (4.2) of
Ref. \cite{Ohashi4})
\begin{eqnarray}
\delta{\hat \rho}({\bf r},\omega)
=
\delta{\hat \rho}_U({\bf r},\omega)
+
\delta{\hat \rho}_{g_{\rm r}}({\bf r},\omega),
\label{eq.3.7}
\end{eqnarray}
where 
\begin{eqnarray}
\delta{\hat \rho}\equiv
\left(
\begin{array}{c}
\delta\rho_1\\
\delta\rho_2\\
\delta\rho_3
\end{array}
\right).
\label{eq.3.7b}
\end{eqnarray}
In Eq. (\ref{eq.3.7}), $\delta{\hat \rho}_U({\bf r},\omega)$ includes the
contribution of non-resonant interaction, which has already appeared
in Eqs. (\ref{eq.2.26})-(\ref{eq.2.28}), given by
\begin{eqnarray}
\delta{\hat \rho}_U({\bf r},\omega)=
-{U \over 2}
\int d{\bf r}'{\hat \Pi}^0({\bf r},{\bf r}',\omega)
\delta{\hat \rho}({\bf r}',\omega).
\label{eq.3.8}
\end{eqnarray}
Here the 
matrix elements of ${\hat \Pi}^0({\bf r},{\bf r}',\omega)$ are given by
Eqs. (\ref{eq.2.19})-(\ref{eq.2.25}), with $\Delta({\bf r})$ being
now replaced with ${\tilde \Delta}({\bf r})$. The effect of Feshbach
resonance appears in $\delta{\hat \rho}_{g_{\rm r}}({\bf r},\omega)$
as
\begin{eqnarray}
\delta{\hat \rho}_{g_{\rm r}}({\bf r},\omega)
&=&
{g_{\rm r}^2 \over 2}
\int d{\bf r}'\int d{\bf r}''
{\hat B}({\bf r},{\bf r}',\omega)
{\hat \Pi}^0({\bf r}',{\bf r}'',\omega)
\delta{\hat \rho}({\bf r}'',\omega).
\label{eq.3.9}
\end{eqnarray}
In contrast to the non-resonance part $U$ in Eq. (\ref{eq.3.8}), the
effective interaction described by 
$g_{\rm r}^2{\hat B}({\bf r},{\bf r}',\omega)$ 
in Eq. (\ref{eq.3.9}) is {\it non-local and
frequency-dependent}, reflecting that it is mediated by Bose
excitations. The interaction kernel ${\hat B}({\bf r},
{\bf r}',\omega)$ is a $3\times 3$-matrix, where only $B_{11}$,
$B_{12}$, $B_{21}$ and $B_{22}$ are finite,
indicating that it only
works in the Cooper-channel. These matrix elements can be
expressed in the form
\begin{eqnarray}
\left(
\begin{array}{cc}
B_{11} & B_{12}  \\
B_{21} & B_{22}  \\
\end{array}
\right)=
{\hat W}{\hat D}_0({\bf r},{\bf r}',\omega){\hat W}^\dagger,
\label{eq.3.10}
\end{eqnarray}
where ${\hat W}$ is the unitary matrix,
\begin{eqnarray}
{\hat W}={1 \over \sqrt{2}}
\left(
\begin{array}{cc}
1 &1 \\
i & -i \\
\end{array}
\right).
\label{eq.3.11}
\end{eqnarray}
${\hat D}_0({\bf r},{\bf r}',\omega)$ is the
matrix single-particle Green's function for a free Bose gas in a
trap\cite{Rykayzen}. Using the eigenfunctions $\Phi_n({\bf r})$ of
${\hat h}_B$ with energy
\begin{equation}
\xi_n=\sum_{\alpha=x,y,z}\Omega^B_\alpha
\Bigl[n_\alpha+{1 \over 2}\Bigr]
+2\nu-2\mu,
\label{eq.3.11b}
\end{equation} 
we can write ${\hat D}_0({\bf r},{\bf r}',\omega)$ as
\begin{eqnarray}
{\hat D}_0({\bf r},{\bf r}',\omega)
=\sum_n{\Phi_n({\bf r})\Phi_n^*({\bf r}') \over 
(\omega+i0^+)\sigma_3-\xi_n}.
\label{eq.3.12}
\end{eqnarray}
\vskip2mm
\subsection{Kohn's theorem in the presence of a Feshbach resonance}
\vskip2mm
In solving the 
coupled equations in Eq. (\ref{eq.3.7}) for a collective mode, it
is useful to note that the contribution from the Feshbach resonance, 
$\delta{\hat \rho}_{g_{\rm r}}({\bf r},\omega)$, 
does not affect the equation for 
$\delta\rho_3({\bf r},\omega)$. 
Thus, we easily find from Eq. (\ref{eq.2.40}) that
\begin{eqnarray}
\left(
\begin{array}{c}
\delta\rho_1({\bf r},\Omega_x) \\
\delta\rho_2({\bf r},\Omega_x) \\
\delta\rho_3({\bf r},\Omega_x)
\end{array}
\right)
=
\left(
\begin{array}{c}
\partial_x{\tilde \Delta}({\bf r}) \\
-2im\Omega_x x{\tilde \Delta}({\bf r}) \\
-{U \over 2}\partial_xn({\bf r})
\end{array}
\right),
\label{eq.3.13}
\end{eqnarray}
is a solution of the equation for $\delta\rho_3$ when $\omega=\Omega_x$.
\par
To see if Eq. (\ref{eq.3.13}) also satisfies the other two equations
for $\delta\rho_1$ and $\delta\rho_2$ in Eq. (\ref{eq.3.8}), we
substitute this {\it trial} solution into the first
($\equiv\delta\rho_{U1}$) and second ($\equiv\delta\rho_{U2}$)
component of Eq. (\ref{eq.3.8}). These can be evaluated in
the same way as in Eqs. (\ref{eq.2.39}) and (\ref{eq.2.41}), and we
obtain at $\omega=\Omega_x$
\begin{eqnarray}
\left(
\begin{array}{c}
\delta\rho_{U1} \\
\delta\rho_{U2}
\end{array}
\right)
&=&
\left(
\begin{array}{c}
\partial_x\Delta({\bf r}) \\
-2im\Omega_xx\Delta({\bf r})
\end{array}
\right)
=
{U \over g_{\rm r}}
\left(
\begin{array}{c}
\partial_x({\hat h}_B\phi_m({\bf r})) \\
-2im\Omega_xx{\hat h}_B\phi_m({\bf r})
\end{array}
\right)
\equiv
{U \over g_{\rm r}}\delta{\bar \rho}_U({\bf r},\Omega_x).
\nonumber
\\
\label{eq.3.14}
\end{eqnarray}
Here we have used the identity in Eq. (\ref{eq.3.5}). 
\par
Next we substitute the trial solution in Eq. (\ref{eq.3.13}) into the
first ($\equiv\delta\rho_{g_{\rm r}1}$) and second
($\equiv\delta\rho_{g_{\rm r}2}$) component of Eq. (\ref{eq.3.9}),
and set $\omega=\Omega_x$. Noting that the factor 
$\int d{\bf r}''{\hat \Pi}_0({\bf r}',{\bf r}'',\Omega_x)
\delta{\hat \rho}({\bf r}'',\Omega_x)$ 
can be simplified by using Eq. (\ref{eq.3.14}), we find
\begin{eqnarray}
\left(
\begin{array}{c}
\delta\rho_{g_{\rm r}1} \\
\delta\rho_{g_{\rm r}2}
\end{array}
\right)
=
-g_{\rm r}\int d{\bf r}'{\hat W}{\hat D}_0
({\bf r},{\bf r}',\Omega_x){\hat W}^\dagger
\delta{\bar \rho}_U({\bf r}',\Omega_x).
\nonumber
\\
\label{eq.3.15}
\end{eqnarray}
Substituting Eqs. (\ref{eq.3.11}), (\ref{eq.3.12}) and (\ref{eq.3.14})
into Eq. (\ref{eq.3.15}), we obtain
\begin{eqnarray}
\delta\rho_{g_{\rm r}1}({\bf r},\Omega_x)
&=&
g_{\rm r}\sum_n
{\Phi_n({\bf r}) \over \xi_n^2-\Omega_x^2}
\int d{\bf r}'\Phi_n^*({\bf r}')
\Bigl[
\xi_n\partial_{x'}({\hat h}_B\phi_m({\bf r}'))
-M\Omega_x^2x'{\hat h}_B\phi_m({\bf r}')
\Bigr],
\nonumber
\\
\label{eq.3.16}
\end{eqnarray}
\begin{eqnarray}
\delta\rho_{g_{\rm r}2}({\bf r},\Omega_x)
&=&
-ig_{\rm r}\Omega_x\sum_n
{\Phi_n({\bf r}) \over \xi_n^2-\Omega_x^2}
\int d{\bf r}'\Phi_n^*({\bf r}')
\Bigl[
M\xi_nx'{\hat h}_B\phi_m({\bf r}')-\partial_{x'}
({\hat h}_B\phi_m({\bf r}'))
\Bigr],
\nonumber
\\
\label{eq.3.17}
\end{eqnarray}
The operation of ${\hat h}_B$ on $\phi_m({\bf r})$ can be conveniently
calculated
by expanding 
the molecular condensate wavefunction, 
\begin{equation}
\phi_m({\bf r})=\sum_n\alpha_n\Phi_n({\bf r}).
\label{eq.3.18}
\end{equation}
Here $\Phi_n({\bf r})$ are the eigenfunctions of a non-interacting
Bose gas of molecules in a trap [see Eq. (\ref{eq.3.12})].
\par
Putting all these together, we have
\begin{eqnarray}
\delta\rho_{g_{\rm r}1}({\bf r},\Omega_x)
&=&
g_{\rm r}\sum_{nn'}\alpha_{n'}\xi_{n'}
{\Phi_n({\bf r}) \over \xi_n^2-\Omega_x^2}
\int d{\bf r}'\Phi_n^*({\bf r}')
\Bigl[
\xi_n\partial_{x'}\Phi_{n'}({\bf r}')-M\Omega_x^2x'\Phi_{n'}({\bf r}')
\Bigr],
\nonumber
\\
\label{eq.3.19}
\end{eqnarray}
\begin{eqnarray}
\delta\rho_{g_{\rm r}2}({\bf r},\Omega_x)
&=&
-ig_{\rm r}\Omega_x\sum_{nn'}\alpha_{n'}\xi_{n'}
{\Phi_n({\bf r}) \over \xi_n^2-\Omega_x^2}
\int d{\bf r}'\Phi_n^*({\bf r}')
\Bigl[
M\xi_nx'\Phi_{n'}({\bf r}')-\partial_{x'}\Phi_{n'}({\bf r}')
\Bigr].
\nonumber
\\
\label{eq.3.20}
\end{eqnarray}
The integration over ${\bf r}'$ can be carried out by expressing $x'$
and $\partial_{x'}$ in terms of raising and lowering operator of a
harmonic oscillator for a molecule, 
\begin{eqnarray}
{\hat a}^\dagger\equiv{1 \over \sqrt{2M\Omega_x^B}}[-\partial_x+M\Omega_x^Bx],
\label{eq.3.21}
\end{eqnarray}
\begin{eqnarray}
{\hat a}\equiv {1 \over \sqrt{2M\Omega_x^B}}[\partial_x+M\Omega_x^Bx].
\label{eq.3.22}
\end{eqnarray}
Since ${\hat a}\Phi_n({\bf r})=\sqrt{n_x}\Phi_{n-1}({\bf r})$ and 
${\hat a}^\dagger\Phi_n({\bf r})=\sqrt{n_x+1}\Phi_{n+1}({\bf r})$ (where
$n\pm1$ is an abbreviation for $(n_x\pm1,n_y,n_z)$), Eqs. (\ref{eq.3.19}) and
(\ref{eq.3.20}) reduce to, respectively,
\begin{eqnarray}
\delta\rho_{g_{\rm r}1}({\bf r},\Omega_x)
&=&
\sqrt{M\Omega_x^B \over 2}g_{\rm r}
\sum_{nn'}\alpha_{n'}\xi_{n'}
{\Phi_n({\bf r}) \over \xi_n^2-\Omega_x^2}
\int d{\bf r}'\Phi_n^*({\bf r}')
\Bigl[
(\xi_n-\Omega_x){\hat a}
-
(\xi_n+\Omega_x){\hat a}^\dagger
\Bigr]
\Phi_{n'}({\bf r}')
\nonumber
\\
&=&
\sqrt{M\Omega_x^B \over 2}g_{\rm r}
\Bigl[
\sum_n
\sqrt{n_x+1}\alpha_{n+1}{\xi_{n+1} \over \xi_n+\Omega_x}\Phi_n({\bf r})
-
\sum_{n~({n_x\ge 1})}
\sqrt{n_x}\alpha_{n-1}{\xi_{n-1} \over \xi_n-\Omega_x}\Phi_n({\bf r})
\Bigr],
\nonumber
\\
\label{eq.3.23}
\end{eqnarray}
\begin{eqnarray}
\delta\rho_{g_{\rm r}2}({\bf r},\Omega_x)
&=&
-i\sqrt{M \over 2\Omega_x^B}\Omega_xg_{\rm r}
\sum_{nn'}\alpha_{n'}\xi_{n'}
{\Phi_n({\bf r}) \over \xi_n^2-\Omega_x^2}
\int d{\bf r}'\Phi_n^*({\bf r}')
\Bigl[
(\xi_n-\Omega_x){\hat a}
+
(\xi_n+\Omega_x){\hat a}^\dagger
\Bigr]
\Phi_{n'}({\bf r}')
\nonumber
\\
&=&
-i\sqrt{M \over 2\Omega_x^B}\Omega_x g_{\rm r}
\Bigl[
\sum_{n}
\sqrt{n_x+1}\alpha_{n+1}{\xi_{n+1} \over \xi_n+\Omega_x}\Phi_n({\bf r})
+
\sum_{n~({n_x\ge 1})}
\sqrt{n_x}\alpha_{n-1}{\xi_{n-1} \over \xi_n-\Omega_x}\Phi_n({\bf r})
\Bigr].
\nonumber
\\
\label{eq.3.24}
\end{eqnarray}
\par
When the trap frequency for molecules $\Omega_x^B$ is equal to
that of Fermi atoms $\Omega_x$, the factors 
$\xi_{n\pm 1}/(\xi_n\pm\Omega_x)$ disappears because of 
$\xi_{n\pm 1}\equiv\xi_n\pm\Omega_x^B=\xi_n\pm\Omega_x$. 
In this particular case, we find
\begin{eqnarray}
\delta\rho_{g_{\rm r}1}({\bf r},\Omega_x)
&=&
\sqrt{M\Omega_x \over 2}g_{\rm r}
\sum_n
\sqrt{n_x+1}\alpha_{n+1}\Phi_n({\bf r})
-
\sqrt{M\Omega_x \over 2}g_{\rm r}
\sum_{n~({n_x\ge 1})}
\sqrt{n_x}\alpha_{n-1}\Phi_n({\bf r}),
\nonumber
\\
&=&
g_{\rm r}\sqrt{M\Omega_x \over 2}[{\hat a}-{\hat a}^\dagger]
\sum_n\alpha_n\Phi_n({\bf r})
\nonumber
\\
&=&
g_{\rm r}\partial_x\phi_m({\bf r}),
\label{eq.3.25}
\end{eqnarray}
\begin{eqnarray}
\delta\rho_{g_{\rm r}2}({\bf r},\Omega_x)
&=&
-i\sqrt{M\Omega_x \over 2} g_{\rm r}
\sum_{n}
\sqrt{n_x+1}\Phi_n({\bf r})
-
i\sqrt{M\Omega_x \over 2} g_{\rm r}
\sum_{n~({n_x\ge 1})}
\sqrt{n_x}\alpha_{n-1}\Phi_n({\bf r})
\nonumber
\\
&=&
-ig_{\rm r}\sqrt{M\Omega_x \over 2}[{\hat a}+{\hat a}^\dagger]
\sum_n\alpha_n\Phi_n({\bf r})
\nonumber
\\
&=&-2img_{\rm r}\Omega_xx\phi_m({\bf r}).
\label{eq.3.26}
\end{eqnarray}
From Eqs. (\ref{eq.3.14}), (\ref{eq.3.25}) and (\ref{eq.3.26}), 
we find that the upper two
components of the RHS of Eq. (\ref{eq.3.7}) are described by the
composite order parameter ${\tilde \Delta}({\bf r})=
\Delta({\bf r})+g_{\rm r}\phi_m({\bf r})$ as
\begin{eqnarray}
\left(
\begin{array}{c}
\delta\rho_{U1}+\delta\rho_{g_{\rm r}1} \\
\delta\rho_{U2}+\delta\rho_{g_{\rm r}2} \\
\end{array}
\right)
=
\left(
\begin{array}{c}
\partial_x{\tilde \Delta}({\bf r}) \\
-2im\Omega_xx{\tilde \Delta}({\bf r})
\end{array}
\right).
\label{eq.3.27}
\end{eqnarray}
Thus we have proven that Eq. (\ref{eq.3.13}) is a solution of the collective
mode equation (\ref{eq.3.7}) with frequency $\omega=\Omega_x$
(in the special case when $\Omega_x^B=\Omega_x$). We can also show that
\begin{eqnarray}
\left(
\begin{array}{c}
\delta\rho_1({\bf r},\Omega_y) \\
\delta\rho_2({\bf r},\Omega_y) \\
\delta\rho_3({\bf r},\Omega_y)
\end{array}
\right)
=
\left(
\begin{array}{c}
\partial_y{\tilde \Delta}({\bf r}) \\
-2im\Omega_y y{\tilde \Delta}({\bf r}) \\
-{U \over 2}\partial_yn({\bf r})
\end{array}
\right),
\label{eq.3.28}
\end{eqnarray}
\begin{eqnarray}
\left(
\begin{array}{c}
\delta\rho_1({\bf r},\Omega_z) \\
\delta\rho_2({\bf r},\Omega_z) \\
\delta\rho_3({\bf r},\Omega_z)
\end{array}
\right)
=
\left(
\begin{array}{c}
\partial_z{\tilde \Delta}({\bf r}) \\
-2im\Omega_z z{\tilde \Delta}({\bf r}) \\
-{U \over 2}\partial_zn({\bf r})
\end{array}
\right),
\label{eq.3.29}
\end{eqnarray}
are also the solutions of Eq. (\ref{eq.3.7}) with frequency
$\omega=\Omega_y$ 
and 
$\Omega_z$, respectively. These explicit solutions describe the 
Kohn mode in the coupled
Fermion-Boson model.
\par
We now briefly discuss the physical meaning of the Kohn mode
in the presence of a Feshbach resonance, 
given by Eq. (\ref{eq.3.13}).
The density oscillation 
(see $\delta\rho_3({\bf r},\Omega_x)$ in Eq.(\ref{eq.3.13}))
is found to have 
the same form as Eq. (\ref{eq.2.41b2}) obtained in the absence
of the Feshbach resonance. 
On the other hand, the oscillation of the
Cooper-pair order parameter, 
\begin{equation}
\Delta({\bf r},t)=\Delta({\bf r}) 
-C{U \over 2}\partial_x{\tilde \Delta}({\bf r})\cos(\Omega_x t)
+iCUmx\Omega_x{\tilde \Delta}({\bf r})\sin(\Omega_x t)
\label{eq.3.29b}
\end{equation}
cannot be written in the form given in 
Eq. (\ref{eq.2.41e}), because the {\it composite order parameter}
appears in the RHS of Eq. (\ref{eq.3.29b}).
This reflects
the fact that the Cooper-pair oscillations are strongly coupled with
molecular Bose excitations through the Feshbach resonance.
Indeed, the linear response of
the Bose condensate $\phi_m({\bf r})$ induced by the 
oscillation of Fermi atoms described by Eq. (\ref{eq.3.13}) 
is given by  
\begin{equation}
\langle\delta\Phi({\bf r},\Omega_x)\rangle=
{g_{\rm r} \over 2}
\int d{\bf r}'D^{0}_{11}({\bf r},{\bf r}',\Omega_x)
[\delta\rho_1({\bf r}',\Omega_x)-i\delta\rho_2({\bf r}',\Omega_x)].
\label{eq.3.30}
\end{equation}
Here $D_{11}^{0}({\bf r},{\bf r}',\Omega_x)$ is the diagonal component of 
molecular Bose Green's function defined in Eq. (\ref{eq.3.11b}).
Substituting Eq. (\ref{eq.3.13})
(which we multiply with the factor $C$) 
and Eq. (\ref{eq.3.11b}) into Eq. (\ref{eq.3.30}),
and using the same method used in 
Eqs. (\ref{eq.3.23})-(\ref{eq.3.26}), we obtain (when $\Omega_x^B=\Omega_x$)
\begin{equation}
\langle\delta \Phi ({\bf r},\Omega_x)\rangle=
-C{g_{\rm r} \over 2}
[\partial_x-M\Omega_xx]{\hat h_B}^{-1}
{\tilde \Delta}({\bf r}),
\label{eq.3.31}
\end{equation}
where the molecular Hamiltonian 
${\hat h}_B$ is defined just before Eq. (\ref{eq.3.6}). In the same way,
we also obtain
\begin{equation} 
\langle\delta\Phi^\dagger ({\bf r},\Omega_x)\rangle=
-C{g_{\rm r} \over 2}
[\partial_x+M\Omega_xx]{\hat h_B}^{-1}
{\tilde \Delta}({\bf r}).
\label{eq.3.31b}
\end{equation}
As a result, the oscillation of
the Bose condensate associated with the Kohn mode is given by
\begin{equation}
\phi_m({\bf r},t)=
\phi_m({\bf r})
-C{g_{\rm r} \over 2}
\partial_x
(
{\hat h}_B^{-1}{\tilde \Delta}({\bf r})
)
\cos(\Omega_x t)
+
iCm\Omega_xxg_{\rm r}{\hat h}_B^{-1}
{\tilde \Delta}({\bf r})\sin(\Omega_xt).
\label{eq.3.32}
\end{equation}
\par
Putting Eqs (\ref{eq.3.29b}) and (\ref{eq.3.32}) together, the
oscillation of the {\it composite} order parameter 
${\tilde \Delta}({\bf r},t)=\Delta({\bf r},t)+g_{\rm r}\phi_m({\bf r},t)$
is given by
\begin{eqnarray}
{\tilde \Delta}({\bf r},t)={\tilde \Delta}({\bf r})
-{C \over 2}\partial_x
\Bigl(
[U+{g_r^2 \over {\hat h}_B}]{\tilde \Delta}({\bf r})
\Bigr)
\cos(\Omega_xt)
+iCm\Omega_xx
[U+{g_r^2 \over {\hat h}_B}]{\tilde \Delta}({\bf r}).
\label{eq.3.33}
\end{eqnarray}
We note that $U+{g_r^2 \over {\hat h}_B}$ describes the interaction
between atoms, where ${g_r^2 \over {\hat h}_B}$ involves the
dynamical effect associated with the Feshbach resonance.
Indeed, in a uniform Fermi gas,
when one neglects the kinetic energy of Bose molecules, this factor
reduces to $U+g_{\rm r}/(2\nu-2\mu)$, which has been previously obtained
as the pairing interaction associated with the Feshbach resonance
\cite{Timmermans2,Holland1,Ohashi1,Ohashi3,Ohashi4}.
Using Eqs. (\ref{eq.3.5}) and (\ref{eq.3.18}), 
we can expand the equilibrium composite order parameter in terms 
of the eigenfunctions of the Bose molecules $\Phi_n({\bf r})$ in a
trap,
\begin{eqnarray}
{\tilde \Delta}({\bf r})
&=&g_{\rm r}\sum_n
\alpha_n
\Bigl(
1+{U \over g_{\rm r}^2}\xi_n
\Bigr)
\Phi_n({\bf r})
\nonumber
\\
&\equiv&
\sum_n\beta_n\Phi_n({\bf r}).
\label{eq.3.34}
\end{eqnarray}
Substituting Eq. (\ref{eq.3.34}) into Eq. (\ref{eq.3.33}), we find
\begin{eqnarray}
{\tilde \Delta}({\bf r},t)
&=&
\sum_n\beta_n\Phi_n({\bf r})
-{C \over 2}\sum_n\beta_n
\Bigl[
U+{g_{\rm r}^2 \over \xi_n}
\Bigr]
\partial_x\Phi_n({\bf r})\cos(\Omega_xt)
\nonumber
\\
&+&
iCm\Omega_xx\sum_n\beta_n
\Bigl[
U+{g_{\rm r}^2 \over \xi_n}
\Bigr]
\Phi_n({\bf r})\sin(\Omega_xt)
\nonumber
\\
&\simeq&
\sum_n\beta_n
\Phi_n
\Bigl(
{\bf r}-{\bf e}_x{C \over 2}[U+{g_{\rm r}^2 \over \xi_n}]
\cos(\Omega_x t)
\Bigr)
e^{iCm\Omega_xx(U+g_{\rm r}^2/\xi_n)\sin(\Omega_xt)}.
\label{eq.3.35}
\end{eqnarray}
Equation (\ref{eq.3.35}) shows that each eigenfunction $\Phi_n({\bf r})$ 
rigidly oscillates around the center of mass with
the frequency $\Omega_x$. However, since the
amplitude of this oscillation in each component 
(which is given by ${C \over 2}[U+{g_{\rm r}^2 \over \xi_n}]$)
is different,
the oscillation of the composite order parameter,
which is given by 
$|{\tilde \Delta}({\bf r},t)|^2$, is not described as
$|{\tilde \Delta}({\bf r}+{\bf e}_xC'\cos(\Omega_xt))|^2$.
This is quite different from the 
Kohn mode in a Fermi superfluid
in the absence of a Feshbach resonance [see Eq.(\ref{eq.2.41e})].
This difference arises because the effective interaction associated with
the Feshbach resonance involves the {\it dynamical} effect of molecular
Bosons, so that the interaction between atoms depends on energy
as $U+{g_{\rm r}^2 \over \xi_n}$. In the BEC limit, where
the composite order parameter is described by the ground state
of the harmonic potential ($\Phi_0({\bf r})$), only the component
with $n=0$ remains in Eq. (\ref{eq.3.35}). In this limiting case,
Eq. (\ref{eq.3.35}) does reduce to  
$|{\tilde \Delta}({\bf r},t)|^2=
|{\tilde \Delta}({\bf r}+{\bf e}_xC'\cos(\Omega_xt))|^2$.
Thus in this limit, we arrive at the usual
Kohn mode solutions for a Bose condensed gas,
involving an oscillation of the equilibrium order parameter.
\par
We might note that the Kohn mode solutions
which appear in the
linear response functions 
describing the Fermi fields
also appear in the
Bose excitation spectrum.
This is to be expected, since in the presence of a 
Feshbach resonance, one knows\cite{Ohashi1,Ohashi3,Ohashi4} 
that the collective modes
of the fermions associated with Cooper-pairs are strongly
coupled to the molecular excitations.
 The $2\times 2$-matrix 
renormalized Bose Green's function ${\hat D}$ 
in the HFB-GRPA is given by\cite{Ohashi4,Ranninger}
\begin{eqnarray}
{\hat D}(\omega)=
\Bigl[
1-{\hat \Sigma}(\omega){\hat D}^0(\omega)
\Bigr]^{-1}
{\hat D}^0(\omega).
\label{eq.3.35b}
\end{eqnarray}
Here we have used matrix notation for the
dependence on ${\bf r}$ [see Eq. (\ref{eq.3.9})].
The self-energy correction ${\hat \Sigma}(\omega)$ includes the effect
of fluctuations in the Fermi atoms, given by
\begin{equation}
{\hat \Sigma}(\omega)={g_{\rm r}^2 \over 2}
{\hat W}^\dagger
{\hat \eta}
\Biggl[
{\hat \Pi}^0(\omega)
\Bigl[
1+{U \over 2}{\hat \Pi}^0(\omega)
\Bigr]^{-1}
\Biggr]
{\hat W},
\label{eq.3.36}
\end{equation}
where the projection operator ${\hat \eta}[{\hat A}]$ extracts the
(11), (12), (21), and (22) components from a $3\times 3$-matrix
${\hat A}$. The molecular Bose excitation spectrum
is determined from the poles of Eq. (\ref{eq.3.35b}), 
given by the condition that the determinant vanishes,
\begin{eqnarray}
0
&=&
det
\Biggl[
1-{g_{\rm r}^2 \over 2}{\hat D}^0(\omega)
{\hat W}^\dagger
{\hat \eta}
\Biggl[
{\hat \Pi}^0(\omega)
\Bigl[
1+{U \over 2}{\hat \Pi}^0(\omega)
\Bigr]^{-1}
\Biggr]
{\hat W}
\Biggr]
\nonumber
\\
&=&
{det
\Biggl[
1+{1 \over 2}
\Bigl[
U-g_{\rm r}^2{\hat B}(\omega)
\Bigr]
{\hat \Pi}^0(\omega)
\Biggr]
\over
det\Bigl[1+{U \over 2}{\hat \Pi}^0(\omega)\Bigr]
}.
\label{eq.3.37}
\end{eqnarray}
In the last expression, the determinant is taken over
$3\times 3$-matrices in the 
$({\hat \rho}_1,{\hat \rho}_2,{\hat \rho}_3)$-space.
Equation (\ref{eq.3.37}) always has solutions
corresponding to the poles of the Fermi linear response functions.
Thus,
the renormalized 
Bose Green's functions in Eq. (\ref{eq.3.35b}) also exhibit
the Kohn mode solutions at $\omega=\Omega_\alpha$ 
(when $\Omega^B_\alpha=\Omega_\alpha$).
\par
As discussed in Sec. II, the extension to include the strong-coupling
effect based on the NSR theory does not destroy our proof. 
Thus, we can safely study collective modes in the BCS-BEC
crossover region by using the NSR-GRPA formalism 
even in the presence of the Feshbach
resonance, without any breakdown of Kohn's theorem, as long as the atomic
and molecular trap frequencies are identical 
($\Omega_\alpha^B=\Omega_\alpha$). 
\par
When the molecules and atoms have different trap frequencies
($\Omega_\alpha^B\ne\Omega_\alpha$), the key solutions in 
Eqs. (\ref{eq.3.25}) and (\ref{eq.3.26})
are no longer satisfied. In this regard, we recall that, in the
presence of a Feshbach resonance, the dominant particles continuously
change from unpaired Fermi atoms to
molecular bosons as one goes through the strong-coupling BEC regime 
(i.e., decrease the threshold energy
$2\nu$\cite{Ohashi1,Ohashi3,Ohashi4,Milstein,Ohashi5}). 
As a result,
the {\it average} trap frequency which the dominant particles feel also
changes from $\Omega_\alpha$ to $\Omega_\alpha^B$, 
which in turn must affect the
frequency of the ``Kohn mode" in the BCS-BEC crossover. The breakdown
of Kohn's theorem when $\Omega_\alpha^B\ne\Omega_\alpha$ is not due to the
approximation used in HFB-GRPA (or NSR-GRPA), but rather is due to the
changing nature of the excitation spectrum 
peculiar to a trapped Fermi gas with a Feshbach
resonance.
\par
Fig.1 shows the calculated 
frequency of the dipole mode at $T=0$ in the BCS-BEC
crossover given by an approximate theory.
 We find that Kohn's theorem holds well (within our
numerical accuracy) when $\Omega_B=\Omega_F$. On the other hand, the
mode frequency {\it depends on} the threshold energy $2\nu$ in the
BCS-BEC crossover when $\Omega_B\ne\Omega_F$. It continuously
changes from the trap frequency of Fermi atoms $\Omega_F$ to that of
Bose molecules $\Omega_B$ as one passes through the BCS-BEC crossover
regime, as one expects.
\vskip2mm
\subsection{Goldstone's theorem in the presence of a Feshbach resonance}
\vskip2mm
In this final subsection, 
we briefly discuss the zero frequency Goldstone mode in the
coupled Fermion-Boson model. At $\omega=0$, the interaction kernel
${\hat B}$ in Eq. (\ref{eq.3.10}) is proportional to the unit matrix,
so that, as in the BCS model discussed in Sec.II C, we need only 
consider the phase fluctuation
component $\delta\rho_2$ in Eq. (\ref{eq.3.7}). This has the form
\begin{eqnarray}
\delta\rho_2({\bf r},0)=
-{U \over 2}
\int d{\bf r}'\Pi^0_{22}({\bf r},{\bf r}',0)\delta\rho_2({\bf r}',0)
+{g_{\rm r}^2 \over 2}
\int d{\bf r}'\int d{\bf r}'' 
D^0_{22}({\bf r},{\bf r}',0)\Pi^0_{22}({\bf r}',{\bf r}'',0).
\nonumber
\\
\label{eq.4.1}
\end{eqnarray}
When we take $\delta\rho_2({\bf r},0)={\tilde \Delta}({\bf r})$, we
find following the discussion in Sec. II.C that the first term in the RHS
of this equation reduces to $\Delta({\bf r})$. Using the same method
used to derive Eq. (\ref{eq.3.15}), we can
integrate over ${\bf r}''$ in the second term of Eq. (\ref{eq.4.1}) 
($\equiv\delta\rho_{g_{\rm r}2}$) to obtain
\begin{eqnarray}
\delta\rho_{g_{\rm r}2}
&=&
-g_{\rm r}
\int d{\bf r}'D^0_{22}({\bf r},{\bf r}',0){\hat h}_B\phi_m({\bf r}').
\label{eq.4.2}
\end{eqnarray}
Using Eqs. (\ref{eq.3.12}) (with $\omega=0$) and (\ref{eq.3.18}), we find
\begin{eqnarray}
\delta\rho_{g_{\rm r}2}
&=&
g_{\rm r}
\sum_{nn'}\alpha_{n'}{\xi_{n'} \over \xi_n}\Phi_n({\bf r})
\int d{\bf r}'
\Phi^*_n({\bf r}')\Phi_{n'}({\bf r}')
\nonumber
\\
&=&
g_{\rm r}\sum_n\alpha_n\Phi_n({\bf r})
\nonumber
\\
&=&
g_{\rm r}\phi_m({\bf r}).
\label{eq.4.3}
\end{eqnarray}
Thus the RHS of Eq. (\ref{eq.4.1}) has been shown explicitly
to equal $\Delta({\bf 
  r})+g_{\rm r}\phi_m({\bf r})\equiv{\tilde \Delta}({\bf r})$. 
This proves that 
$\delta\rho_2({\bf r},0)={\tilde \Delta}({\bf r})$ describes the 
zero frequency Goldstone
mode. In contrast to the BCS model in the absence of a
Feshbach resonance, however, the Goldstone mode is now a 
collective phase oscillation of the {\it composite} order parameter, 
including
contributions associated with 
both Cooper-pairs and the molecular condensate. We conclude that
 the HFB-GRPA (and NSR-GRPA) formalism is consistent with
 Goldstone's theorem even in the presence of the Feshbach resonance.
We note that Goldstone mode arises even if the trap frequencies felt by
Fermi atoms and Bose molecules are different.
We also note that our result also guarantees that 
the renormalized Bose Green's function
in Eq. (\ref{eq.3.35b}) has a gapless (zero frequency) excitation,
a required condition for any approximate
theory used to study Bose condensation.
\vskip3mm
\section{Summary}
\par
In this paper, we have proved that the HFB-GRPA is a consistent
formalism with Kohn's theorem at all temperatures. This proof is
also valid for the strong-coupling superfluid theory developed 
by Nozi\`eres and
Schmitt-Rink, as used in Ref.\cite{Ohashi4}. 
Using the NSR-GRPA formalism, we can safely study the linear response
dynamics of the superfluid phase in the BCS-BEC
crossover without breakdown of this general theorem on the dipole
oscillation. 
The relevance of the Kohn mode in BCS superfluids was never discussed much
in the context of superconductivity. In the case of superfluid
Fermi gases trapped
in a parabolic potential, it is important that any
approximate theory used to calculate collective modes be
consistent with Kohn's theorem.
\par
We also considered, for the first time, 
the effect of a Feshbach resonance and the associated
formation of molecules on the Kohn mode. When the molecules feel
the same trap frequency as that for Fermi atoms, we explicitly proved that
HFB-GRPA and NSR-GRPA lead to Kohn's theorem. However, when
the molecular trap frequency is different from the atomic trap
frequency (which can arise when dealing with different hyperfine
states), 
the dipole mode
frequency depends on the strength of the effective
interaction (through $2\nu$) associated with the Feshbach resonance 
in the BCS-BEC
crossover region. 
This was to be expected, of course, since the dominant excitations
continuously change from Fermi atoms to Bose molecules as we go through
the crossover regime (i.e., decrease the base molecular threshold $2\nu$). 
This result is a clear experimental signature of different
trap frequencies. 
\par
We also have given, for the first time,
 a detailed discussion of the various quantities
which are oscillating in the Kohn mode in a trapped superfluid Fermi
gas, as given by our explicit results in Eqs. (\ref{eq.2.34}) and 
(\ref{eq.3.13}).
In particular, we showed that with a Feshbach resonance,
the Kohn mode involves a much more complex oscillation 
[see discussion after Eq. (\ref{eq.3.29b})]
than without
a Feshbach resonance [see discussion after Eq. (\ref{eq.2.41b})].
\par
The BCS-BEC crossover and the Feshbach resonance are key phenomena in
current studies on superfluidity in ultra-cold Fermi gases, such as
$^{40}$K and $^6$Li. In considering these phenomena, one should be
careful to introduce approximations which do not break Kohn's
theorem, which is an exact result of a many-body system in a harmonic
trap. Our proof on the consistency of HFB-GRPA and
NSR-GRPA with both Kohn's theorem and Goldstone's theorem 
shows that they can be used to study
the collective modes of
{\it strongly-correlated} superfluid Fermi gases.
Results of such calculations will be reported elsewhere\cite{Ohashizz}. 
\par
\vskip2mm
\acknowledgments
I would like to thank Prof. Allan Griffin for many useful discussions 
and comments, as well as a  
critical
reading of the manuscript. 
This work was financially supported by a Grant-in-Aid for Scientific 
research
from the 
Ministry of Education, Culture, Sports, Science and Technology of Japan,
as well as from a University of Tsukuba Research Project. 
\par
%

%
\newpage
\begin{figure}
\includegraphics[width=7cm,height=5cm]{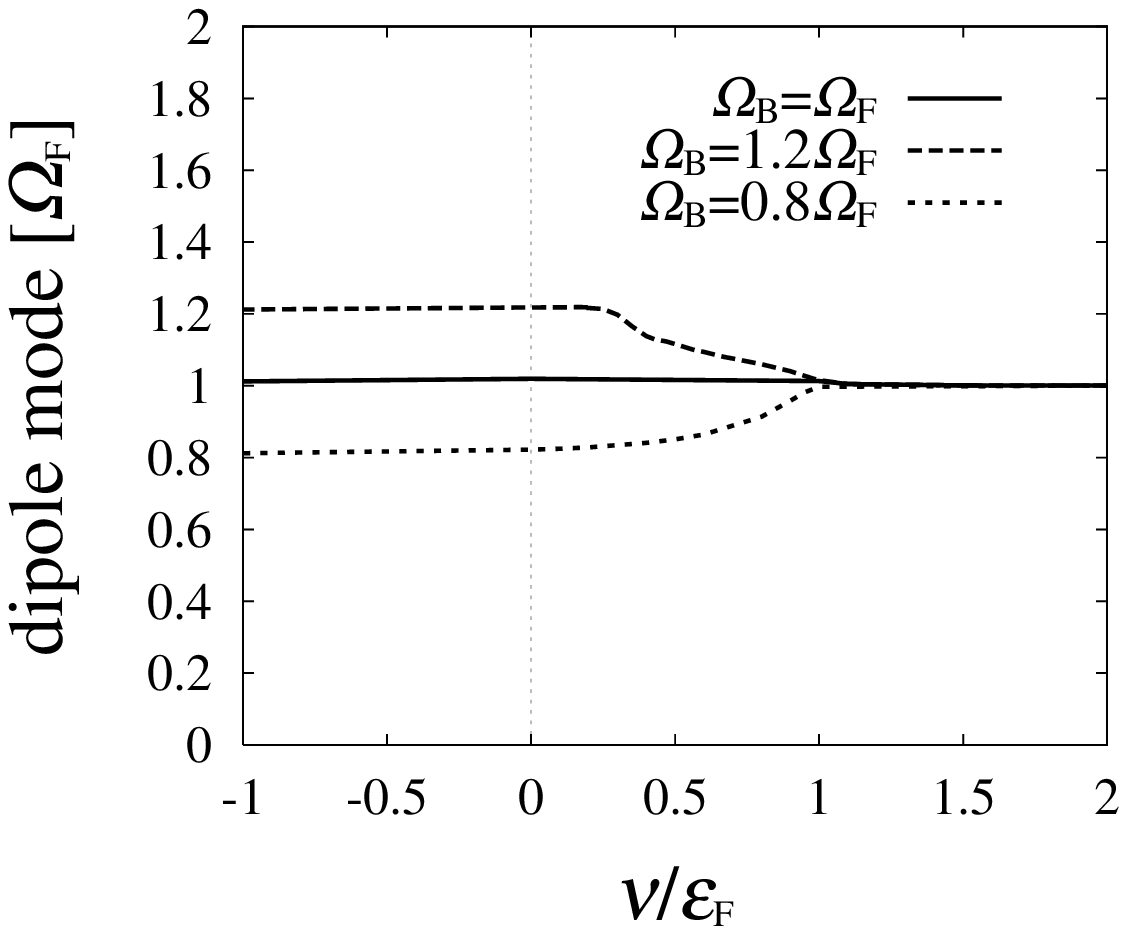}%
\caption{\label{fig1} 
Calculated frequency of the dipole mode at $T=0$ 
in the superfluid phase of a trapped
Fermi gas with a Feshbach resonance. We consider an isotropic harmonic
trap with $\Omega_\alpha=\equiv\Omega_F$ and
$\Omega_\alpha^B\equiv\Omega_B$. The Fermi energy
$\varepsilon_{\rm F}=31.5\Omega_F$ is for a free Fermi gas, with
$N=10,912$ atoms. We take $U=0.001\Omega_F$,
$g_{\rm r}=0.06\Omega_F$ (which gives 
$UN/R_{\rm F}^3=0.35\varepsilon_{\rm F}$ and
$g_{\rm r}\sqrt{N}/R_{\rm F}^{3/2}=0.2\varepsilon_{\rm F}$,
where $R_{\rm F}\equiv\sqrt{2\varepsilon_{\rm F}/m\Omega_F^2}$ 
is the Thomas-Fermi radius), and
$\omega_c=161.5\Omega_F$. 
}
\end{figure}
\par
%
%
%

\begin{thebibliography}{99}
\bibitem{note2} When we consider pseudo-spin fluctuations, one
 should also include the interaction in the spin-channel given by 
$H_{\rm SF}\equiv{U \over 4}\int d{\bf r}s_z({\bf r})s_z({\bf r})$, 
where $s_z({\bf r})=\sum_\sigma\sigma\Psi_\sigma^\dagger({\bf r})
\Psi_\sigma({\bf r})$ is a spin density operator. Since the 
spin-fluctuations do not couple with either the density fluctuations or 
the superfluid fluctuations, we do not consider $H_{\rm SF}$ in this paper.
\end{thebibliography}
\end{document}